\documentclass[lettersize,journal]{IEEEtran}

\ifCLASSINFOpdf
\usepackage{amsmath,amsfonts}
\usepackage{algorithmic}
\usepackage{algorithm}
\usepackage{array}
\usepackage[caption=false,font=footnotesize,labelfont=rm,textfont=rm]{subfig}
\usepackage{textcomp}
\usepackage{stfloats}
\usepackage{url}
\usepackage{verbatim}
\usepackage{graphicx}
\usepackage{multirow}
\usepackage{cite}
\begin{document}

\title{SLICE: {\bf S}LO-Driven Scheduling for {\bf L}LM {\bf I}nferen{\bf c}e on {\bf E}dge Computing Devices}

\author{\IEEEauthorblockN{Will Chow}
}

\markboth{Journal of \LaTeX\ Class Files,~Vol.~14, No.~8, August~2025}%
{Shell \MakeLowercase{\textit{et al.}}: A Sample Article Using IEEEtran.cls for IEEE Journals}

%\IEEEpubid{\copyright~IEEE}

\maketitle

\begin{abstract}
Large Language Models (LLMs), as the foundational architecture for next-generation interactive AI applications, not only power intelligent dialogue systems but also drive the evolution of embodied intelligence on edge devices, including humanoid robots, smart vehicles, and other scenarios. The applications running on these edge devices impose differentiated Service Level Objectives (SLO) requirements on LLM services, specifically manifested as distinct constraints on Time to First Token (TTFT) and Time Per Output Token (TPOT) as well as end-to-end latency. Notably, edge devices typically handle real-time tasks that are extremely sensitive to latency, such as machine control and navigation planning. However, existing scheduling service systems still prioritize maximizing output token throughput as the sole optimization objective, failing to adequately address the diversity of SLO requirements. This ultimately results in persistently high violation rates for end-to-end latency or TPOT related SLOs.

This paper proposes SLICE, an innovative scheduling solution designed for edge computing scenarios with differentiated SLO requirements. By combining a utility-maximizing request scheduling algorithm with a dynamic iterative control mechanism for generation rates, SLICE significantly improves LLM inference service SLO attainment. Experimental results demonstrate that compared to state-of-the-art solutions Orca and FastServe, SLICE achieves up to 35× higher SLO attainment and 3.4× advantage in task completion time than the other two solutions. This version is temporarily hosted anonymously for double-blind review.
 
\end{abstract}


\section{Introduction}
\IEEEPARstart{L}{arge} language models (LLMs) such as GPT and DeepSeek series \cite{achiam2023gpt}\cite{liu2024deepseek} are raising artificial general intelligence to an unprecedented heights.  With their human-like generation capabilities, these models are being rapidly adopted across diverse domains, most notably in edge intelligence. This development has sparked new possibilities for intelligent edge devices and is poised to fundamentally transform lifestyles and work patterns. A prominent example is Google's SayCan \cite{ahn2022can}, which significantly enhances robots' ability to comprehend and execute linguistic commands. Leading technology companies including Boston Dynamics \cite{macdonald2024language}, Tesla \cite{cui2023drivellm}, and Apple \cite{alizadeh2024llm} are actively developing and implementing LLM technologies in their products. As a result, integrating the vast knowledge capabilities of large language models into edge devices—spanning smartphones, IoT systems, and automotive platforms—represents a transformative opportunity in modern edge computing ecosystems\cite{cui2024survey}\cite{liu2024edge}\cite{chiu2025v2v}.

\begin{figure}[!t]
	\centering
	\subfloat[Decode latency VS. batch size]{\includegraphics[width=0.47\linewidth]{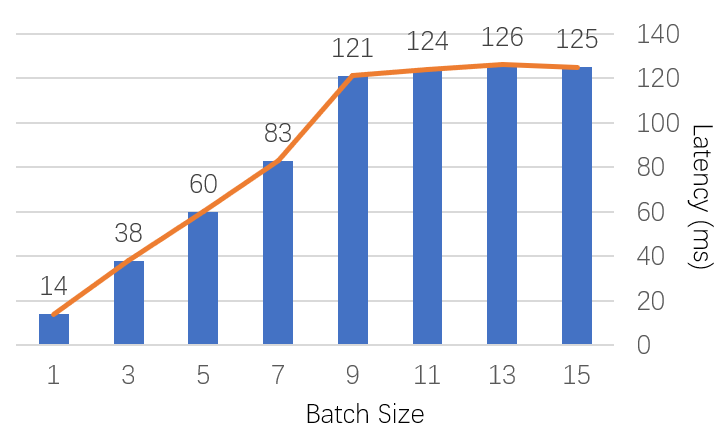}%
		\label{fig_batch_latency_a}}
	\hfil
	\subfloat[Token throughput VS. batch size]{\includegraphics[width=0.47\linewidth]{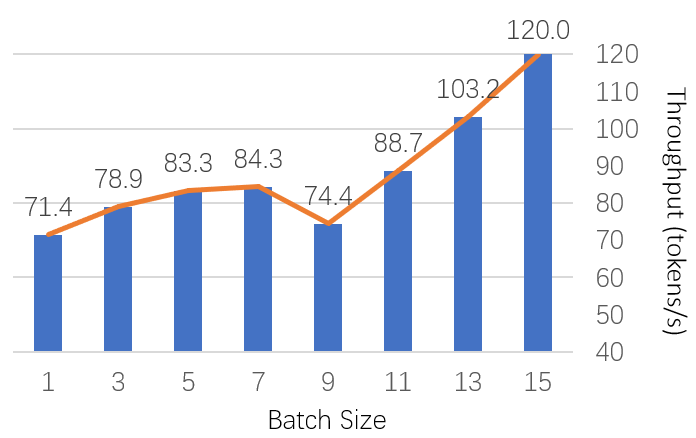}%
		\label{fig_batch_latency_b}}
	\caption{We conducted a batch processing performance evaluation of the ChatGLM2-6B model on an NVIDIA RTX 4060 Ti 16GB GPU. Experimental results demonstrate that as the batch size increases from 1 to 9, the per-token latency shows near-linear growth while throughput improves only marginally. When batch size exceeds 9, although the per-token latency stabilizes and throughput continues to scale linearly with batch size, the absolute latency spikes above 120ms, reducing the token generation rate per task to below 10 tokens/s. This performance bottleneck will significantly impact latency-sensitive real-time tasks, making it impossible to meet strict SLO.}
	\label{fig_batch_latency}
\end{figure}

Deploying large language models on edge devices requires supporting diversified application services, where different applications exhibit significant variations in service-level objectives (SLOs) for inference performance. Real-time applications like navigation planning and device control demand strict end-to-end deadline constraints \cite{ling2024timelyllm,ding2023task,chen2023typefly}. Non-real-time interactive applications focus on Time To First Token (TTFT) and Time Per Output Token (TPOT)\cite{borui2025efficient,hongsola,huang2025slo}. For example, voice interaction must guarantee output speed not less than human speech rate \cite{rayner1998eye}, while text-based question answering requires matching reading speed to maintain interaction fluency\cite{yoo2019comparison}\cite{mctear2024rise}. Given the constrained hardware resources of edge devices and the large scale of LLMs, even a single-unit increase in batch size can lead to significant single-token latency spikes (as illustrated in Fig. \ref{fig_batch_latency}). This tension between resource constraints and multi-SLO requirements places scheduling system design as the key enabler for efficient LLM inference in edge computing scenarios.

\begin{figure}[!t]
	\centering
	\subfloat[Existing coarse-grained decoding scheduling]{\includegraphics[width=0.99\linewidth]{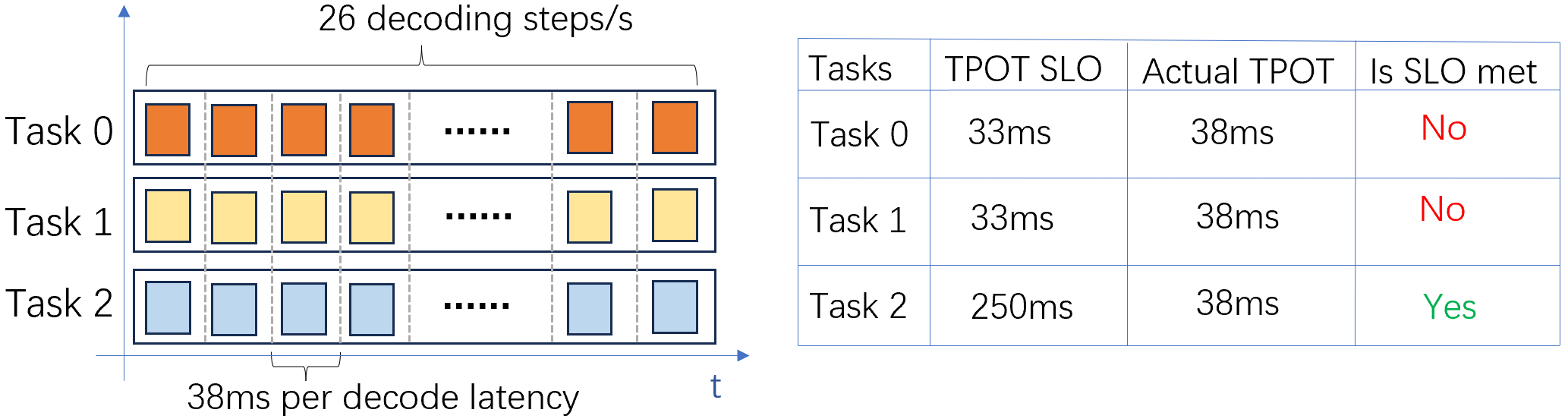}%
		\label{fig_sloaware_example_a}}
	\vfil
	\subfloat[SLO-aware decoding scheduling]{\includegraphics[width=0.99\linewidth]{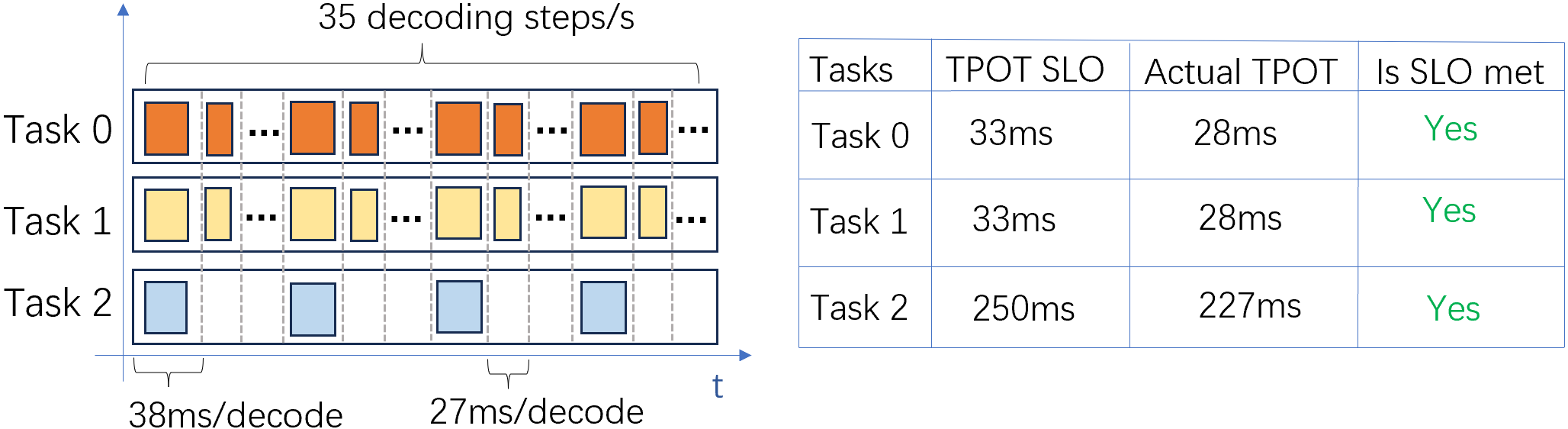}%
		\label{fig_sloaware_example_b}}
	\caption{(a) The current scheduling system cannot perform fine-grained, decode-level scheduling. Each decoding process batches all scheduled tasks together, which increases decoding latency and consequently leads to violations of task-level SLOs. (b) By adopting decode-level batch scheduling and implementing an on-demand allocation approach, more tasks can meet their SLO requirements.}
	\label{fig_sloaware_example}
\end{figure}

Mainstream inference services such as Orca\cite{yu2022orca}, vLLM\cite{kwon2023efficient}, FastLLM\cite{fastllmgithub_repo}, and FasterTransformer \cite{FasterTransformer_repo} predominantly employ First-Come-First-Served (FCFS) scheduling to compose a processing batch, which may cause queuing delay for later-arriving tasks under ultra heavy workloads in cloud computing scieneros. Meanwhile, alternative high-performance schedulers adopt specialized techniques to alleviate this queuing delay. Sarathi Serve \cite{agrawal2024taming} implements chunked pre-filling to eliminate stalls between prefill and decode phase transitions. FastServe \cite{wu2023fast} deploys multi-level feedback queues with iteration-level preemption to mitigate head-of-line blocking, while Splitwise/DistServe \cite{patel2024splitwise} isolates pre-filling and decoding stages for independent optimization. Llumnix \cite{sun2024llumnix} further enhances throughput through load balancing, prioritization, defragmentation, and auto-scaling. All these designs indiscriminately batch tasks regardless of their SLO requirements during inference. During each decoding iteration, all scheduled tasks are indiscriminately batched together and processed through the LLM's decoding forward pass. This coarse-grained decoding scheduling works well for homogeneous SLO scenarios without strict deadline constraints or TPOT requirements, while our experiments reveal its inadequate when handling edge computing workloads with various strict SLO requirements, as illustrated in Fig. \ref{fig_sloaware_example_a}. This experimental case demonstrates that existing decoding scheduling can cause a long decoding latency that exceeds TPOT requirements when the batch size surpasses the critical threshold, resulting in severe SLO violations. More critically, if the deadline of real-time tasks cannot be guaranteed, it will directly impair the normal operation of edge devices, such as those performing delay-sensitive and mission-critical tasks like machine control and navigation planning, potentially leading to system malfunctions or even task failures.

\emph{Motivated by the above insights,  can we design an intelligent scheduling system that dynamically adjusts decoding rates according to tasks' SLO requirements, thereby improving overall SLO attainment while ensuring effective guarantees for real-time tasks' deadline in edge computing scenarios?} %However, all existing systems indiscriminately batch tasks regardless of their SLO requirements during inference, making them inherently unsuitable for multi-SLO scenarios in edge computing. This uniform treatment leads to both elevated SLO violation rates and inefficient hardware utilization, as demonstrated in Figure 2A.

In this paper, we propose a SLO-driven scheduling strategy called SLICE for LLM inference services. The core concept of SLICE revolves around "real-time task prioritization and on-demand resource allocation".  Specifically, SLICE gives priority to scheduling real-time tasks to ensure they meet their deadlines. Additionally, it allocates decoding rates in accordance with SLO requirements. This approach prevents low-demand tasks from excessively consuming computational resources, thereby enabling more concurrent tasks to run even with limited computing power (as shown in Fig. \ref{fig_sloaware_example_b}). By adhering to this principle, SLICE achieves a synergistic improvement in both SLO attainment and the successful completion of real-time tasks within their deadlines. To implement this scheduling concept, SLICE employs a two-stage scheduling strategy.

%The key idea of SLICE is "real-time task prioritization and resource-on-demand allocation". Specifically, SLICE prioritizes the scheduling of real-time tasks to meet their deadlines, and allocates decoding rates based on SLO requirements to prevent low-demand tasks from excessively occupying computational resources and allow more concurrent tasks with limited computing power (as illustrated in Fig. \ref{fig_sloaware_example_b}). Based on this idea, SLICE achieves synergistic improvement in both SLO attainment and real-time tasks' deadline achievement. To realize the scheduling idea, SLICE implements a two-stage scheduling strategy.

%By adhering to the principle of on-demand allocation, SLICE maximizes SLO attainment while ensuring the deadline constraints of real-time tasks. To be specific, SLICE implements a two-stage SLO-aware scheduling strategy. The first stage employs a utility maximization approach for task batch selection, prioritizing real-time task scheduling while maximizing the throughput of non-real-time tasks. The second stage dynamically adjusts decoding rates via a decode-mask matrix, allocating fewer resources to tasks with lower demands under strict SLO constraints, thereby enabling limited resources to support more concurrent tasks.

In the first phase (i.e., the task selection phase), each newly arrived task is assigned a utility value that reflects its importance. Notably, real-time tasks are assigned significantly higher utility values than non-real-time tasks (typically 10 to 100 times higher) in the edge scenarios. By maximizing the total utility value, the system selects a batch of tasks for decoding, ensuring real-time task prioritization while maximizing non-real-time task throughput. 

In the second phase (i.e., the rate allocation phase), the decoding rate is dynamically adjusted based on the SLO requirements of each task. To minimize system overhead, this paper proposes an innovative decode-mask matrix that enables cycle-based scheduling to achieve rate control for individual tasks. Specifically, the number of columns in the matrix represents the total number of decoding operations per cycle, while each row corresponds to one scheduled task, recording its decoding schedule within the cycle. In the matrix, setting the $n$th column of a row to "1" indicates that the corresponding task participates in the $n$th decoding operation of that cycle. For tasks with stricter TPOT SLO requirements (i.e., those requiring a higher decoding rate), the "1"s are distributed more densely across their respective rows, thereby enabling a higher decoding rate. During the actual scheduling process, the system performs column-wise scanning starting from the left to right column in each cycle, triggering a decoding operation upon completing the scan of a column. For every scan, the system groups all tasks corresponding to rows marked "1" in the current column into a single decoding batch. Through this iterative scanning and dynamic batch aggregation mechanism, the system can dynamically adjust the decoding rate of each task according to its SLO requirements.

%The second phase (i.e., the rate allocation phase) dynamically adjusts decoding rates based on each task's SLO requirements. To minimize system overhead, this paper proposes an innovative token-mask matrix for rate control. Specifically, the matrix's columns represent total decoding operations per cycle, while rows strictly correspond to scheduled tasks, recording each task's decoding timeline. The system initiates column-by-column scanning from the first column each cycle, triggering a decoding operation upon scanning each column - all tasks in rows marked '1' in the current column form the decoding batch. Through this iterative scanning and batching process, the system dynamically adjusts task decoding frequencies to meet SLO requirements.

Overall, the main contributions of this paper are as follows:
\begin{itemize}
	\item{This paper investigates the distinctive challenges in LLM task scheduling for edge computing scenarios, where different tasks exhibit significant variations in SLOs for inference performance.}
	\item{The paper proposes a utility-maximizing task selection mechanism that optimizes task scheduling upon arrival, achieving dual objectives of real-time task deadline assurance and maximum overall SLO attainment.}
	\item{The paper proposes a decode-mask matrix periodic scanning mechanism that accurately controls the decoding rate of scheduled tasks even under various load conditions.}
	\item{The paper design and implement SLICE, a SLO-aware LLM inference scheduling strategy built on top of the above utility-maximizing based task selection and decode-mask matrix based rate allocation mechanisms. We evaluate SLICE on various scenarios and demonstrate that it achieves up to 35× improvement in SLO attainment under heavy workloads and a 3.4× advantage in task completion time compared to state-of-the-art solutions Orca \cite{yu2022orca} and FastServe \cite{wu2023fast}.}
\end{itemize}

The rest of this paper is organized as follows. Section \ref{sec_background} presents the research background and related work. Section \ref{sec_motivation} and \ref{sec_slice} provide the motivations and design of SLICE, respectively. Section \ref{sec_evaluation} presents the validation of SLICE, and Section \ref{sec_conclusion} concludes the research contributions.

\section{Background and Related Work} \label{sec_background}
\subsection{Background}
{\bf Applications of LLM in edge computing.}
Edge scenarios include a wide range of devices, including mobile phones, smart homes, vehicles, and robots, which are ubiquitous in users' daily lives \cite{liu2024edge,yao2024minicpm}. Deploying LLM on these edge devices can achieve a wider range of usage, better computing efficiency, more robust offline behavior, and better privacy/security protection, providing a promising solution for more practical edge intelligence applications \cite{ahn2022can,li2025mobillm,macdonald2024language}. For example, in the field of embodied intelligence, as one of the core goals of artificial general intelligence, embodied intelligence constructs self centered intelligent systems through LLM, which can interact with the surrounding environment and have perception, reasoning, and planning capabilities \cite{chen2023typefly,ding2023task}. The emerging large-scale visual language models allow embedded artificial intelligence agents to solve related tasks in a highly end-to-end manner, such as embedded question answering and visual language navigation, voice control, chatbots, artificial intelligence assistants, and so on \cite{wu2023tidybot,zhangvision,cui2024survey}. Therefore, LLM is becoming the foundation model for various applications in edge scenarios, providing context specific inputs (also known as "prompts") to support the inference needs of these applications. However, different applications have different service requirements. Some applications have strong real-time performance and require the inference system to complete inference within a specified time limit, such as navigation planning \cite{chen2023typefly}, industrial robot control \cite{wu2023tidybot,ling2024timelyllm}, etc. Non-real-time tasks prioritize token generation rates or TPOT demands. For instance, voice conversations must sustain a minimum of ‌8 tokens per second‌ to match natural speech flow, whereas text-based question answering requires ‌at least 10 tokens per second‌ to ensure reading fluency \cite{rayner1998eye,yoo2019comparison,mctear2024rise}. This heterogeneous demand introduces distinct challenges for task scheduling on edge devices.

\begin{figure}[!t]
	\centering
	\includegraphics[width=0.99\linewidth]{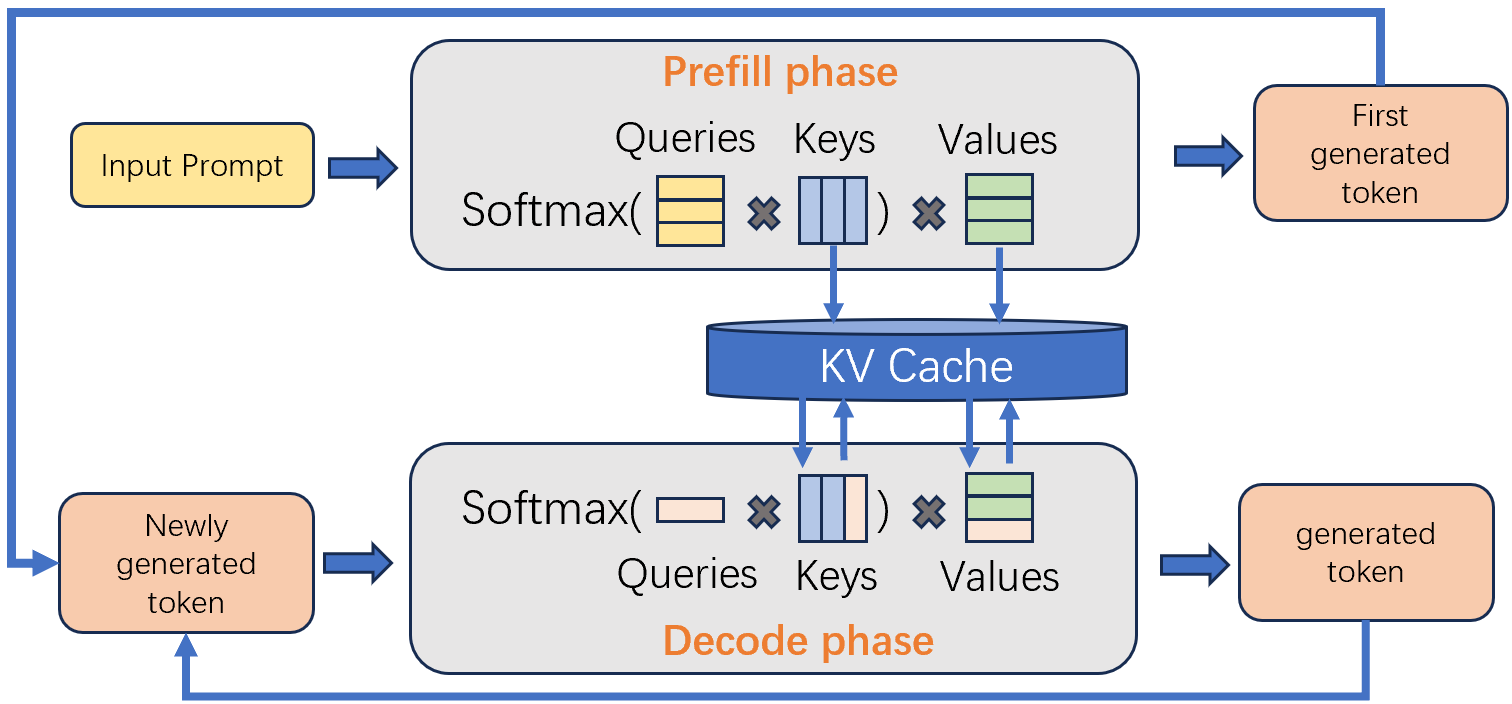}
	\caption{In LLM inference, the prefill and decode phases utilize a dynamically maintained Key-Value (KV) cache. This cache acts as a working memory, being continuously stored, loaded, and updated throughout the generation process.}
	\label{fig_llm_inference}
\end{figure}

{\bf Autoregressive Generation of LLM inference.} LLM belongs to a language model series with Transformer architecture as its core, and its inference mechanism follows the autoregressive paradigm \cite{devlin2019bert,achiam2023gpt}. Under this framework, input consists of a series of tokens, commonly referred to as prompts. LLM parses these prompts and outputs a probability distribution indicating the tokens that may be sampled next. We formally define an inference iteration as the complete cycle from token processing to next-token sampling. Through intensive training on large corpora, LLM is able to perform various language tasks excellently. For example, when inputting "Piece of", the probability of the model predicting "cake" as a subsequent word is significantly higher than "sugar". After the first iteration is completed, the newly generated token will be appended to the end of the initial prompt and input back into LLM as a whole to continue generating the next token, as illustrated in Fig. \ref{fig_llm_inference}. This process repeats until a unique token representing the end of the sequence is generated or the preset maximum output length threshold is reached.

\subsection{Related Work}
Due to GPUs and other accelerators having numerous parallel computing units, they typically batch inference tasks to improve hardware utilization and system throughput. When batching is enabled, input tensors from multiple tasks are concatenated and fed into the model as a single input \cite{FasterTransformer_repo,fastllmgithub_repo}. The drawback of batching is that it incurs higher memory overhead compared to processing individual tasks. Since activation memory scales proportionally with model size, the maximum batch size for LLM inference is constrained by the model's size. Currently, optimization efforts for inference primarily focus on two directions: memory management and batch scheduling optimization.‌

{\bf In memory management,} fairseq \cite{ott2019fairseq} proposes storing keys and values in a key-value cache for reuse throughout the iterative process. This divides inference into two phases: the prefilling phase (initial iteration), where prompts are processed to generate key-value caches for each transformer layer in LLM, and the decoding phase, where LLM only computes queries, keys, and values for newly generated tokens while reusing and updating the cache, resulting in typically shorter iteration times than the first iteration. Other transformer-optimized libraries like HuggingFace \cite{jain2022hugging} and FasterTransformer \cite{FasterTransformer_repo} implement similar optimizations. However, unpredictable KV cache memory demands require systems to reserve maximum GPU capacity for batch tasks, severely limiting batch size benefits. For instance, LLaMA-2-13B\cite{kumar2024towards} requires ~3.2GB of KV cache per request at 4k sequence lengths, competing with model weights (26GB) on mainstream GPUs with only tens of GBs. Recent solutions like vLLM's dynamic memory allocation employ PagedAttention: KV data is stored in dynamically allocated memory blocks as cache scales, with active requests assigned based on available blocks \cite{kwon2023efficient}. Methods like FlexGen \cite{sheng2023flexgen} and CARASERVE \cite{li2024caraserve} further offload partial tasks to CPUs to increase batch size and throughput. These memory optimizations complement the present work.

{\bf In batch scheduling,} early approaches mimicked traditional deep learning batch processing by executing all tasks in a batch before moving to the next, causing completed tasks to remain idle and new arrivals to wait. Orca's iteration-level scheduling \cite{yu2022orca} addressed this by running a single iteration per batch, allowing finished jobs to exit and new ones to join after each token generation. SARATHI \cite{agrawal2024taming} further improved GPU utilization by splitting prompts into chunks for parallel execution with decoding. Splitwise \cite{patel2024splitwise} decoupled prefilling and decoding batches to reduce interference. However, these methods may cause head-of-line blocking due to First-Come-First-Serve (FCFS) scheduling strategy. FastServe \cite{wu2023fast} mitigates this with multi-level feedback queues and distribute tasks to different priority queues according to tasks lengths. vLLM-LTR \cite{fu2024efficient} predicts task length via a machine learning method and executes a Shortest-Job-First (SJF) scheduling. Llumnix \cite{sun2024llumnix} prioritized real-time tasks by migrating instances across GPU pools, but this approach proved infeasible for resource-constrained edge computing.

\section{ Motivation and challenges} \label{sec_motivation}
\subsection{Opportunity: SLO-Aware Scheduling}
The above state-of-the-art scheduling strategies exhibit a critical flaw by failing to adequately address the diverse SLOs of tasks. They adopted a uniform batching mechanism that forcibly groups all scheduled tasks into a single batch to perform the forward computation for each decode stage. This coarse-grained scheduling leads to two contradictory phenomena. On one hand, as the batch size increases, the per-decode latency grows linearly, causing high-generation-rate tasks to fail meeting TPOT requirements and real-time tasks' deadlines due to decoding delays. On the other hand, low-generation-rate tasks are forced to perform decoding at frequencies far exceeding their SLO requirements, resulting in the dilemma of insufficient supply for high-demand tasks and excessive supply for low-demand tasks (see Fig. \ref{fig_sloaware_example} for typical cases). To address this, we propose dynamic rate allocation for individual tasks based on their SLO demands to improve SLO attainment. However, implementing such SLO-aware rate allocation faces two major challenges: {\bf Dual Optimization Objectives} and {\bf Scheduling Overhead}.

\subsection{Challenges}
{\bf Challenge 1: dual optimization objectives.} In practical task scheduling, there is a significant conflict between ensuring the deadline SLO of real-time tasks and maximizing the overall SLO attainment of all tasks. This conflict stems from the stringent deadline requirements of real-time tasks. They need to complete processing within an extremely short time frame, which directly translates into a rigid demand for a high token generation rate. To achieve this, the system must limit the decoding batch size. Otherwise, it cannot meet the real-time tasks' delay requirements for TPOT. As shown in Fig. \ref{fig_batch_latency_a}, blindly increasing the decoding batch size will lead to higher decoding latency, directly causing real-time tasks to fail to meet their predefined SLOs. In contrast, from the perspective of overall task scheduling efficiency, maximizing the SLO achievement rate requires enlarging the decoding batch size as much as possible to enhance the system's parallel processing capability, thereby scheduling more tasks. Therefore, the key challenge in scheduling design lies in finding the optimal balance between ensuring real-time task SLOs and improving the overall SLO attainment.

%The core challenge of batch task scheduling lies in balancing resource utilization and service-level objectives (SLO). When the number of concurrent tasks exceeds the system’s processing capacity, a batch scheduling strategy must be adopted, and determining the appropriate batch size is particularly critical: too small a batch may lead to idle GPU resources and reduced utilization, while too large a batch may increase the average decoding latency, thereby exacerbating the risk of SLO violations. In edge computing scenarios, this issue becomes even more complex, as the system must simultaneously meet the strict latency requirements of real-time tasks and the SLO compliance rate of non-real-time tasks. 
%To address this multi-objective optimization challenge, this paper innovatively proposes a profit-maximizing scheduling framework. This approach quantifies task value by assigning a dynamic profit value to each task (with real-time tasks having significantly higher profit values than non-real-time tasks) and calculates a profit rate based on the profit value and SLO rate requirement. During scheduling, the system prioritizes selecting the ​​b​​ tasks with the highest profit rates to form a batch. Coupled with an intelligent rate allocation mechanism, it ensures that all tasks within the batch meet their SLO constraints while optimizing the overall system profit. This method not only addresses the efficiency of resource allocation but also accommodates the differentiated needs of tasks with varying priorities.

{\bf Challenge 2: scheduling overhead.} Precise rate allocation requires tracking the state of scheduled tasks, which imposes significant overhead. For any task, enforcing a strict fixed time interval for token generation is impractical. This mechanism not only necessitates real-time monitoring of each task’s state but also requires precisely scheduling the generation time for every token. Such an approach leads to two major issues. First, the system incurs substantial scheduling overhead. Second, even if precise generation times are defined, token generation may still be blocked due to hardware resource unavailability, ultimately causing noticeable stuttering. %To address this, our paper employs an iterative matrix-based approach. In each iteration, the matrix selects tasks for batch processing, thereby controlling token generation rates while reducing system load.

\section{SLICE: SLO-Driven Schduling}\label{sec_slice}

To address the aforementioned challenges, this chapter proposes a systematic solution. Section \ref{sec_slice_formulation} constructs a utility maximization model to transform the dual-objective optimization scheduling problem into a mathematical optimization problem, identifying the optimal trade-off between real-time guarantees and overall SLO attainment optimization by maximizing system-wide comprehensive utilities. Section \ref{sec_slice_offline} introduces an offline SLICE scheduling strategy that systematically addresses the aforementioned optimization challenges through its two core components—task scheduling selection and task rate allocation. Section \ref{sec_task_selection} provides a task selection scheme based on a utility-rate-priority decision framework, enabling maximizing system-wide comprehensive utility to resolve Challenge 1. Section \ref{sec_rate_allocation} innovatively proposes an SLO-aware decoding mask matrix to allocate the decoding rates of scheduled tasks, tackling Challenge 2. Section \ref{subsec:off2online} further realizes the effective transformation of the offline SLICE into an online strategy, ensuring efficient and stable execution in real-time scenarios.

\subsection{Problem Formulation}\label{sec_slice_formulation}
We first formulate the problem using the notations in Table \ref{tab:notations}. For ease of description, we translate the deadline constraints of real-time tasks into dual-metric requirements for TTFT and TPOT. Through this transformation, both real-time and non-real-time tasks can be uniformly modeled under consistent constraint conditions, thereby enhancing the standardization and operational feasibility of the modeling approach. And different types of tasks vary in importance or priority. For instance, real-time tasks related to machine control typically have higher priority. To ensure that high-priority tasks are scheduled first, each task is assigned a utility value. The higher the value, the higher the priority. Thus, our optimization objective is to maximize the system's total utility, as shown in Eq. (\ref{eq:maxu0}), where $U_i$ represents the utility value obtained by task $i$, and $X_i$ is a binary decision variable. $X_i=1$ if task i meets its SLO, otherwise $X_i=0$, as defined in Eq. (\ref{eq:maxu1}).
%table
\begin{table}[htp]
	\centering
	\caption{Summary of notations}
	\begin{tabular}{l|p{0.38\textwidth}}
		\hline
		Notation & Description\\
		\hline
		$N$&The number of tasks\\
		$X_i$&A binary decision variable \\
		$l(b)$&Decoding latency with batch size $b$\\
		$U_i$&Utility value of task $i$\\
		$T_{TPOP}^{i}$&Time Per Output Token (TPOT) of task $i$\\
		$T_{TTFT}^{i}$&Time to First Token (TTFT) of task $i$\\
		$t_{TTFT}^{i}$&The actual TTFT of task $i$\\
		$v_i$&The required generation rate of task $i$, defined as $1/T_{TPOP}^{i}$\\
	    $\tilde{v}_i$&Task $i$'s actual generation rate allocated by the system \\
		$T_{period}$&The estimated duration of the scheduling cycle\\
		\hline
	\end{tabular}
	\label{tab:notations}
	\vspace{-1ex}
\end{table}

% formulation
\begin{figure}[htp]
	\centering 
	\begin{equation} 
		\quad\max \quad \sum_{i=0}^{N} X_{i} U_{i} \label{eq:maxu0}
	\end{equation}
	\begin{flushleft}
		\qquad \qquad s.t.
		\begin{equation}
			X_{i}= \left\{ \begin{aligned}
				&1\quad if \: \tilde{v}_i \geq \frac{1}{T_{TPOT}^{i}}\:\: and \:\: t_{TTFT}^{i}\geq T_{TTFT}^{i}\\
				&0\qquad  others
			\end{aligned}
             \right. \label{eq:maxu1}
		\end{equation}
		\begin{equation}
			b=\sum_{i=1}^{N} I(\tilde{v}_i) \label{eq:maxu2}
		\end{equation}
		\begin{equation}
			I(\tilde{v}_i)= \left\{ \begin{aligned}
				&1\quad if \: \tilde{v}_i > 0\\
				&0\qquad  others
			\end{aligned}
			\right.\label{eq:maxu3}
		\end{equation}
		\begin{equation}
		\sum_{i=1}^{N}\tilde{v}_i \leq \frac{b}{l(b)} \label{eq:maxu4}
		\end{equation}
	\end{flushleft}
\end{figure}

The number of scheduled tasks in the system is denoted by $b$. Each scheduled task is assigned a token generation rate $\tilde{v}_i >0$, as specified in Eq. (\ref{eq:maxu2}) and (\ref{eq:maxu3}). The sum of the rates allocated to all scheduled tasks must not exceed the maximum processing capacity of the GPU, as expressed in Eq. (\ref{eq:maxu4}). Here, $l(b)$ denotes the latency for decoding a batch of $b$ tasks, and $b/l(b)$ represents the maximum system throughput when the batch size is $b$. Generally, $l(b)$ is a non-linear function (Fig. \ref{fig_batch_latency_a} provides an example). If $l(b)$ were a linear function, the problem could be reduced to the knapsack problem, which is NP-hard. Therefore, the problem described by Eq. (\ref{eq:maxu0}) to (\ref{eq:maxu4}) is more complex than the knapsack problem and is also NP-hard.

\subsection{SLO-Driven offline Scheduling} \label{sec_slice_offline}
To efficiently solve the above formulated optimization problem, we propose the SLICE algorithm, a two-phase solution comprising task selection and rate allocation. For clarity, we first present the algorithm in an offline scenario where all tasks arrive simultaneously, and subsequently extend it to the online scenario (detailed in subsection \ref{subsec:off2online}) that accommodates dynamic task arrivals and departures.

\begin{algorithm}[htp]
	\caption{SLO-Driven offline Scheduling: SLICE-offline}\label{alg:SLICEoffline}
	\begin{algorithmic}[1]
		\STATE \textbf{Input:} Task set $\mathbb{N}=\{0,1,2,...,N\}$; TPOT value set $\mathbb{T}=\{T_{TPOT}^{0},T_{TPOT}^{1},...,T_{TPOT}^{N}\}$; Utility value set $\mathbb{U}=\{U_0,U_1,...,U_N\}$; For any task $i\in \mathbb{N}$: TPOT requirement $T_{TPOT}^{i}$, Utility value $U_i$ ; Token buffer $tokenBuf$ for streaming response; Event queue $eventQ$
		\STATE \textbf{Output:} Tasks not yet completed
		\STATE $eventQ\gets\emptyset$
		\\$\triangleright$ First step: task selection
		\STATE $\mathbf{b},\mathbb{N}\gets$ \textsc{TaskSelection} $(\mathbb{N},\mathbb{T},\mathbb{U})$ \label{offline_selection}
		\\$\triangleright$ Second step: rate allocation
		%\STATE Start thread \textsc{RateAllocation}$(\mathbf{b},tokenBuf,eventQ )$
		%\STATE \textbf{while} \textsc{RateAllocation} is not completion \textbf{do}
		%\STATE \hspace{0.5cm}\textbf{if} $tokenBuf\neq\emptyset$ \textbf{do}
		%\STATE \hspace{0.5cm}\hspace{0.5cm} streaming response tokens in $tokenBuf$
		%\STATE \hspace{0.5cm}\textbf{end if}
		%\STATE \textbf{end while}
		\STATE $\mathbf{b}\gets$ \textsc{RateAllocation}$(\mathbf{b},tokenBuf,eventQ )$ \label{offline_rateallocation}
		\STATE $\mathbb{N}\gets\mathbf{b}\cup\mathbb{N}$
		\STATE \textbf{return} $\mathbb{N}$\quad$\triangleright$ Return tasks not yet completed 
	\end{algorithmic}
	\label{alg_SLICE}
\end{algorithm}

In the offline scenario, all tasks arrive simultaneously at time 0. The SLICE scheduling algorithm completes task scheduling through a two-phase mechanism of \textbf{task selection} and \textbf{ rate allocation}, as depicted in Algorithm \ref{alg:SLICEoffline}. The SLICE scheduling algorithm first employs a task selection algorithm to select a batch of tasks (line \ref{offline_selection}), ensuring that the SLOs of these tasks are met while maximizing the overall system utility. The selected batch is then handed over to the rate allocation algorithm for executing the decoding operations of LLMs (line \ref{offline_rateallocation}). This algorithm takes into full consideration the unique decoding computational characteristics of LLMs and dynamically allocates corresponding decoding rates based on the differentiated SLO requirements of each task. It strictly guarantees that the actual decoding rate assigned to each task is no lower than its preset SLO requirement, thereby ensuring service quality while achieving precise scheduling of computing resources.

The detailed design of the task selection algorithm and the rate allocation algorithm can be found in following Section \ref{sec_task_selection} and Section \ref{sec_rate_allocation}.

\subsection{Task Selection} \label{sec_task_selection}
{\bf The task selection process consists of two steps:} utility rate calculation and task selection, as shown in Algorithm \ref{alg:task_select}. 

\textbf{Step 1: Utility Rate Calculation}  (line \ref{task_select_ratecal}). Since the system has a limited number of tokens that can be generated per second, we calculate the utility rate $r_i$ for each task according to Eq. (\ref{eq:rate}). This rate represents the utility gained per token generated within a second. If the token interval of task $i$ is $T_{TPOT}^{i}$, the number of tokens generated per second is $1/T_{TPOT}^{i}$. By distributing the task utility $U_i$ evenly across all tokens generated per second, the utility value per token is given by $\frac{X_i}{1/T_{TPOT}^{i}}=X_i \cdot T_{TPOT}^{i} $, which simplifies to Eq. (\ref{eq:rate}):
\begin{equation}
	r_i = U_i\cdot T_{TPOT}^{i} \label{eq:rate} 
\end{equation}

\textbf{Step 2: Task Selection} (line \ref{task_select_tasksel}-\ref{task_select_end}). In this step, the $|\mathbf{b}|$ tasks with the highest utility rate are selected to scheduling. However, the determination of $\mathbf{b}$ is non-trivial. If too many tasks are selected, it may not be possible to allocate sufficient generation rates to each task due to hardware limitations. Conversely, selecting too few tasks may underutilize hardware resources, resulting in an insufficient number of schedulable tasks. 

To deal with this challenge, this scheme employs a heuristic scheduling strategy. It adopts a non-replacement iterative strategy for task selection. In each iteration, it identifies the task with the highest utility rate from the remaining task set, adds it to the selected task set $\mathbf{b}$ to form the current candidate batch (line \ref{task_selection_selecri}-\ref{task_selection_cupb}), and then evaluates whether the system processing capacity can meet the SLO requirements of this batch (line \ref{task_select_evaslo}). If the requirements can be met, the tasks selected in the current iteration are validated and retained in set $\mathbf{b}$, proceeding to the next iteration; if not, the current iteration's tasks are removed from set $\mathbf{b}$ and returned to the pending task pool, immediately terminating the iteration process (line \ref{task_select_ifslo}-\ref{task_select_endifslo}). The final task set in set $\mathbf{b}$ is determined as the optimal scheduling result that satisfies the SLO constraints.

It should be particularly noted that the rate allocation mechanism (see section \ref{sec_rate_allocation}) of SLICE adopts a periodic scheduling strategy where each scheduling cycle generates at least $v_i$ tokens for each scheduled task $i\in \mathbf{b}$, and when the duration of a single cycle does not exceed 1000ms, the system can ensure that the SLO requirements of all tasks are met. Conversely, if the cycle duration exceeds 1000 milliseconds, the SLO requirements cannot be satisfied (see line \ref{task_select_ifslo}).

The above two steps ensure that the total rate of scheduled tasks strictly adheres to the system’s capacity constraints while prioritizing the execution of tasks with high utility rates, thereby maximizing the total utility of the scheduling sytem defined by Eq. (\ref{eq:maxu0}).
\begin{algorithm}[htp]
	\caption{\textsc{TaskSelection}}\label{alg:task_select}
	\begin{algorithmic}[1]
		\STATE \textbf{Input:} Task set $\mathbb{N}=\{0,1,2,...,N\}$; for any task $i\in \mathbb{N}$: TPOT requirement $T_{TPOT}^{i}$, Utility value $U_i$ 
		\STATE \textbf{Output:} Selected tasks $\mathbf{b}=\{\sigma(1),\sigma(2),...,\sigma(\|\mathbf{b}\|)\}$
		\STATE$\mathbf{b} \gets \emptyset$; \quad $\triangleright$ Initial empty batch
		\STATE$rateSum \gets 0$;
		\\$\triangleright$ Step 1: Calculating utility rate
		\STATE$ r_i \gets  U_i \cdot T_{TPOT}^{i} $ \textbf{ for } $ i\in \mathbb{N} $; \label{task_select_ratecal}
		\\$\triangleright$ Step 2: Select the $|\mathbf{b}|$ tasks with the highest utility rate 
		\STATE\textbf{for} $ i = 0,1,...,N $ \textbf{ do } \label{task_select_tasksel}
		\STATE\hspace{0.5cm}$ \sigma(i) \gets \arg \max_{j\in\mathbb{N}} r_j$;\label{task_selection_selecri}
		%\STATE\hspace{0.5cm}$rateSum\gets rateSum + 1/T_{TPOT}^{\sigma(i)}$;
		\STATE\hspace{0.5cm}$latencySum\gets0$
		\STATE\hspace{0.5cm}$\mathbf{b}\gets \mathbf{b}\cup \sigma(i)$\label{task_selection_cupb}
		\STATE\hspace{0.5cm}$\mathbb{N} \gets \mathbb{N}\setminus \sigma(i)$
		\STATE\hspace{0.5cm}$\mathbf{b}\gets sortTasksBySLORateDescending(\mathbf{b})$
		\\\hspace{0.5cm}$\triangleright$ Estimating the duration for a period
		\STATE\hspace{0.5cm}$period\gets v_{b[i]}l(i+1)+ \sum_{j=0}^{i-1}(v_{b[j]}-v_{b[j+1]})l(j+1)$\label{task_select_evaslo}
		
		%\STATE\hspace{0.5cm}\textbf{for} $j=0,1,...,i$ \textbf{do}
		%\STATE\hspace{0.5cm}\hspace{0.5cm}\textbf{if} $j<i$ \textbf{do}
		%\STATE\hspace{0.5cm}\hspace{0.5cm}\hspace{0.5cm} $period\gets period+(v_{b[j]}-v_{b[j+1]})l(j+1)$
		%\STATE\hspace{0.5cm}\hspace{0.5cm}\textbf{else}
		%\STATE\hspace{0.5cm}\hspace{0.5cm}\hspace{0.5cm} $period\gets period+v_{b[j]}l(j+1)$
		%\STATE\hspace{0.5cm}\hspace{0.5cm}\textbf{end if}
		%\STATE\hspace{0.5cm}\textbf{end for} 
		\STATE\hspace{0.5cm}\textbf{if} $period \geq 1000ms$ \textbf{then}\label{task_select_ifslo}
		\STATE\hspace{0.5cm}\hspace{0.5cm}$\mathbf{b}\gets \mathbf{b}\setminus \sigma(i)$
		\STATE\hspace{0.5cm}\hspace{0.5cm}$\mathbb{N} \gets \mathbb{N}\cup \sigma(i)$
		\STATE\hspace{0.5cm}\hspace{0.5cm}\textbf{break};
		\STATE\hspace{0.5cm}\textbf{end if}\label{task_select_endifslo}
		\STATE\textbf{end for}\label{task_select_end}
		\STATE\textbf{return}  $\mathbf{b},\mathbb{N}$
	\end{algorithmic}
	\label{alg_taskselect}
\end{algorithm}

\subsection{Rate Allocation}\label{sec_rate_allocation}
\begin{figure}[h]
	\centering
	\includegraphics[width=0.99\linewidth]{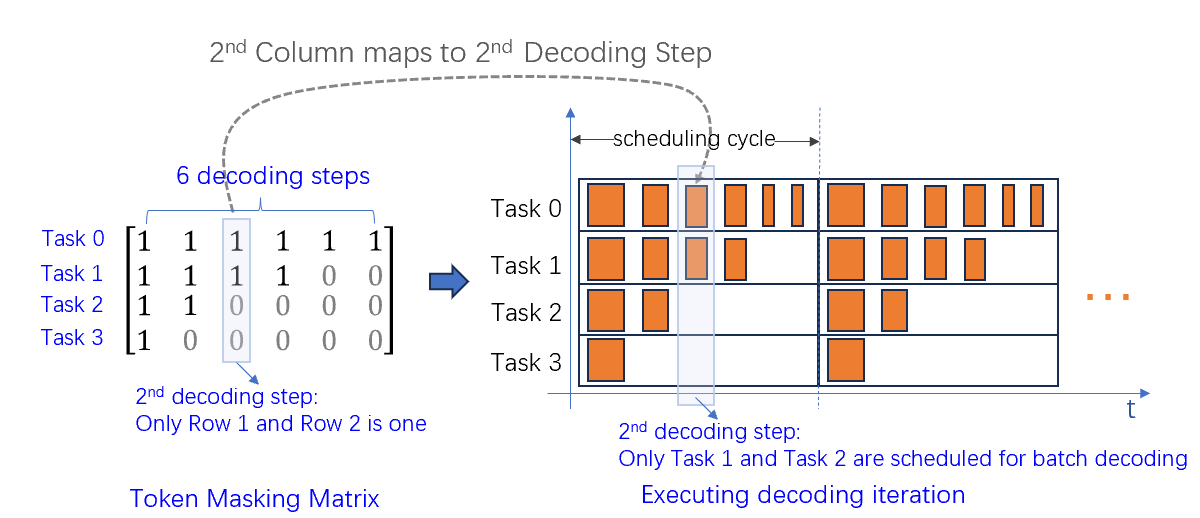}
	\caption{Rate allocation example. The rate allocation algorithm achieves differentiated rate allocation by constructing a 4×6 token generation mask matrix. The system first sorts the four tasks (task0-task3) in descending order of their decoding rate requirements (6/4/2/1 tokens/s) and builds a 4×6 binary scheduling matrix: the row index corresponds to the task ID, where the 0th row (all 1s) ensures task0 is scheduled 6 times per cycle, the 1st row (first 4 bits as 1s) enables task1 to be scheduled 4 times per cycle, the 2nd row (first 2 bits as 1s) realizes task2 being scheduled 2 times per cycle, and the 3rd row (first bit as 1) meets task3's requirement of 1 scheduling per cycle. During execution, a periodic column-wise scanning mechanism is adopted—each scheduling cycle sequentially scans columns 0 to 5, dynamically grouping tasks with a value of 1 in the current column for batch decoding (e.g., when scanning column 2, task0 and task1 corresponding to the 0th and 1st rows are grouped). This matrix-based scheduling approach not only guarantees that each task meets its SLO-required decoding rate but also enables efficient utilization of system resources.}
	\label{fig_rateallocationexample}
\end{figure}
%To ensure the fluency of token generation, the number of tokens generated during task execution must satisfy the following constraint: at any moment from the start to the end of a task, the cumulative number of tokens generated by the system shall be no less than the number required for application output, as specified in Eq. (\ref{eq:raterq}):
%\begin{equation}
%	\int_{0}^{t} v_{i}(t) \leq \frac{t}{T_{TPOT}^{i}}, \quad t\in [0,T_i] %\label{eq:raterq} 
%\end{equation}
%For example, assuming that the human natural speech rate corresponds to a token generation rate of 4 tokens per second, if the system generates 8 tokens in the first second, then even if it pauses generation in the second second, the tokens accumulated in the first second can still meet the output demand. This ensures the continuity of speech output and avoids stuttering.

The rate allocation algorithm employs a periodic scheduling mechanism to achieve precise token generation control. Each scheduling cycle allocates a token generation quota of $vi$ (defined as $1/T_{TPOT}^{i}$) for every scheduled task $i$, with the duration of each cycle strictly limited to within 1000 milliseconds to ensure all tasks' SLOs are rigorously met. Within each cycle, multiple iterations of scheduling are performed, where each iteration selectively groups only a subset of task set $\mathbf{b}$ into dynamic batches for decoding. Tasks with stricter TPOT requirements (i.e., demanding higher token generation rates) receive proportionally higher iteration frequencies, and vice versa. This sophisticated integration of TPOT-weighted probabilistic selection and dynamic batch reorganization techniques ultimately achieves fine-grained SLO-aware rate allocation, optimizing system throughput efficiency while guaranteeing quality of service, as presented in Algorithm \ref{alg:rate_Allocation}.

\textbf{The algorithm can be summarized into two steps:} decode-mask matrix construction and SLO-aware decoding.

\textbf{Step 1: Decode-mask mask matrix construction} (line \ref{ratealloc_sortb}-\ref{ratealloc_matrix_end}). We first sort the selected $|\mathbf{b}|$ tasks in descending order based on their token generation rate requirements and construct a token generation mask matrix to achieve periodic scheduling. Specifically, for each task $i$, a binary mask vector of length $v_0$ is constructed, where the first $v_i$ elements are set to 1 and the remaining elements to 0. The mask vectors of the $|\mathbf{b}|$ tasks are stacked according to the same sorting order, forming a matrix of dimension $|\mathbf{b}|\times v_1$, with the row order strictly corresponding to the task sorting order.

As illustrated in Fig. \ref{fig_rateallocationexample}, there are four tasks requiring decoding execution, which are ranked by their decoding rate requirements in descending order as task0 through task3. The decoding rate requirements for task0 to task3 are 6, 4, 2, and 1 tokens per second, respectively. Accordingly, SLICE constructs a 4×6 matrix: the 0th row consists entirely of 1s (six 1s), indicating that task0 is scheduled for all six decoding steps in each cycle; the 1st row has the first four elements as 1s and the remaining elements as 0s, corresponding to task1 being scheduled for the first four decoding steps in each cycle; the 2nd row has the first two elements as 1s and the remaining elements as 0s, indicating that task2 is scheduled for the first two decoding steps in each cycle; the 3rd row has only the first element as 1 and the remaining elements as 0s, corresponding to task3 being scheduled for the first decoding operation in each cycle.

\begin{algorithm}[htp]
	\caption{\textsc{RateAllocation}}\label{alg:rate_Allocation}
	\begin{algorithmic}[1]
		\STATE \textbf{Input:} A batch of task $\mathbf{b}$; Output token buffer $tokenBuf$; An interruption event queue $eventQ$ used to receive decoding interrupt instructions
		\STATE \textbf{Output:} Token generation rate $v_1,v_2,...,v_T$
		\\$\triangleright$ Construct token masking matrix
		\STATE$ \mathbf{b} \gets sortTasksBySLORateDescending(\mathbf{b})$;\label{ratealloc_sortb}
		\STATE$v_0 \gets \lceil 1/T_{TPOT}^{\mathbf{b}[0]}\rceil$;
		\STATE$ \mathbf{M} \gets [0]_{|\mathbf{b}|\times v_0}$;
		\STATE\textbf{for} $k=0,1,2,...,|\mathbf{b}|\!\!-\!\!1$ \textbf{do}\label{ratealloc_matrix_begin}
		\STATE\hspace{0.5cm} $v_k \gets \lfloor 1/T_{TPOT}^{\mathbf{b}[k]}\rfloor$;
		\STATE\hspace{0.5cm} \textbf{for} $m=0,1,2,...,v_k$ \textbf{do}
		\STATE\hspace{0.5cm}\hspace{0.5cm} $\mathbf{M}[k][m]\gets 1$
		\STATE\hspace{0.5cm} \textbf{end for}
		\STATE\textbf{end for}\label{ratealloc_matrix_end}
		\\$\triangleright$ Decoding execution
		\STATE\textbf{while} True \textbf{do} \label{ratealloc_exestart}
		\STATE\hspace{0.5cm}\textbf{for} $j=0,1,2,...,v_0$ \textbf{do}
		\STATE\hspace{0.5cm}\hspace{0.5cm}$\mathbf{m}\gets \mathbf{M}[j]$\quad$\triangleright$get the $j$th column
		\STATE\hspace{0.5cm}\hspace{0.5cm}$\mathbf{c}\gets \emptyset$
		\STATE\hspace{0.5cm}\hspace{0.5cm}\textbf{for }$k=0,1,2,...,|\mathbf{b}|\!\!-\!\!1$ \textbf{do}\label{ratealloc_batchbegin}
		\STATE\hspace{0.5cm}\hspace{0.5cm}\hspace{0.5cm}$\mathbf{c}\gets \mathbf{c}\cup \mathbf{b}[k]$ \textbf{if} $\mathbf{m}[k]=1$;
		\STATE\hspace{0.5cm}\hspace{0.5cm}\textbf{end for}
		\STATE\hspace{0.5cm}\hspace{0.5cm}$outputTokens\gets excuteBatchForward(\mathbf{c})$\label{ratealloc_batchend}
		\STATE\hspace{0.5cm}\hspace{0.5cm}$\mathbf{e}\gets \emptyset$
		\STATE\hspace{0.5cm}\hspace{0.5cm}\textbf{if} an ending token in $outputTokens$ \textbf{then}
		\STATE\hspace{0.5cm}\hspace{0.5cm}\hspace{0.5cm}$\mathbf{e}\gets$find the ending tasks
		\STATE\hspace{0.5cm}\hspace{0.5cm}\textbf{end if}
		\STATE\hspace{0.5cm}\hspace{0.5cm}$\mathbf{b}\gets \mathbf{b}\setminus\mathbf{e}$\quad $\triangleright$remove the ending tasks
		\\\hspace{0.5cm}\hspace{0.5cm}$\triangleright$restore the output tokens
	    \STATE\hspace{0.5cm}\hspace{0.5cm}$outputBuf\gets outputBuf\cup outputTokens$
	    \STATE\hspace{0.5cm}\hspace{0.5cm}\textbf{if} $\mathbf{b}$ is empty \textbf{then}
	    \STATE\hspace{0.5cm}\hspace{0.5cm}\hspace{0.5cm}\textbf{return} $\emptyset$
	    \STATE\hspace{0.5cm}\hspace{0.5cm}\textbf{end if}
		\STATE\hspace{0.5cm}\hspace{0.5cm}\textbf{if} $eventQ$ is not empty \textbf{then}
		\STATE\hspace{0.5cm}\hspace{0.5cm}\hspace{0.5cm}\textbf{return} $\mathbf{b}$
		\STATE\hspace{0.5cm}\hspace{0.5cm}\textbf{end if}
		\STATE\hspace{0.5cm}\textbf{end for}
		\STATE\textbf{end while}  \label{ratealloc_exeend}
		%\STATE \hspace{0.5cm}\textbf{return}  $\textsc{sign}( \mathbf{H} \beta )$
	\end{algorithmic}
\end{algorithm}

\textbf{Step 2: SLO-aware decoding} (line \ref{ratealloc_exestart}-\ref{ratealloc_exeend}). During decoding execution, a column-wise scanning mechanism is adopted. The system sequentially scans the matrix columns from left to right, moving to the next column after completing the scanning and decoding of the current column. During the scanning of a column, all tasks corresponding to the elements with a value of 1 in that column are grouped into a batch for executing a decoding step (line \ref{ratealloc_batchbegin}-\ref{ratealloc_batchend}). When the scanning and decoding of the rightmost column is completed, a full scheduling cycle is finished, and the system automatically resets to the leftmost column to begin a new cycle. Through this cyclic column-scanning mechanism, the system can accurately achieve dynamic allocation of differentiated rates for each task. The algorithm will continue execution until either all tasks are completed or an interruption event occurs.

As shown in Fig. \ref{fig_rateallocationexample}, during the decoding execution phase, the SLICE algorithm performs periodic scheduling through column-wise scanning. Each scheduling cycle scans sequentially from the 0th column to the 5th column, after which the next cycle restarts scanning from the 0th column and repeats cyclically. During each scan, the system groups tasks corresponding to the positions with a value of 1 in the current column into a batch for a decoding step. For example, when scanning the 2nd column, if the 0th and 1st elements in that column are 1, the SLICE algorithm will group task0 and task1 into a batch and perform one decoding step.

\textbf{Latency Estimation.} Notably, the number of batched tasks in each decoding step may vary, which implies that the decoding latency for each step could also differ. Specifically, a smaller number of tasks in a batch corresponds to lower latency. We can estimate the total latency of a scheduling cycle through the token generation mask matrix, with the specific formula provided below:
\begin{equation}
	T_{period} = v_{b}l(b+1)+ \sum_{j=0}^{b-1}(v_{j}-v_{j+1})l(j+1) \label{eq:period_estimate} 
\end{equation}
where a set of tasks are indexed by from $0$ to $b$, and thus $b+1$ represents the number of scheduled tasks requiring decoding rate allocation. Here, term $v_{b}l(b+1)$ captures the latency for executing $v_{b}$ decoding steps with a batch size of $b+1$, which corresponds to processing the first $v_{b}$ columns in the mask matrix. And the expression $(v_{j}-v_{j+1})l(j+1)$ represents the latency for performing $(v_{j}-v_{j+1})$ decoding steps with a batch size of $j+1$, corresponding to processing columns from $v_{j+1}$ to $v_{j}$ in the mask matrix.

Since the task selection algorithm in Section \ref{sec_task_selection} has accurately estimated the scheduling cycle and selected $b$ tasks through this formula Eq. (\ref{eq:period_estimate}), the scheduling cycle for executing these $b$ tasks via the rate allocation algorithm will not exceed 1000 milliseconds, thereby ensuring that the actual token generation rate obtained by each task is no less than its SLO requirement.

\subsection{Offline to Online consideration}\label{subsec:off2online}
%In this subsection, we generalize the algorithm to the online scenario considering the dynamic arrival and departure of tasks.
In this subsection, we generalize the algorithm to the online scenario considering the dynamic arrival and departure of tasks. To achieve this extension, we adopt a strategy of cyclically executing the offline SLICE algorithm, thereby transforming it into an online SLICE algorithm, as depicted in Algorithm \ref{alg:SLICE_online}. During each execution of the offline SLICE algorithm, the scheduling system continuously monitors task arrivals and departures. If any task arrival or departure is detected, the system immediately interrupts the decoding phase of the offline SLICE algorithm and restarts its execution process (line \ref{task_exe_event_begin}-\ref{task_exe_event_end}). Specifically, upon restarting, if new tasks have arrived, they are incorporated into the current task set to participate in subsequent task selection. Those selected will proceed to the scheduling phase and initiate decoding execution, while those not selected remain unscheduled. If task departures are detected, the remaining task set is input into the offline SLICE algorithm to re-execute the task selection and decoding processes, ensuring that the scheduling results consistently reflect the latest task status.

\textbf{Preemption Consideration.} To enhance the algorithm's adaptability to diverse scenarios, the online algorithm introduces a dynamic utility value adjustment mechanism after the decoding phase (line \ref{task_utility_adapt}), thereby enabling flexible customization of preemption strategies. For example, by reducing the utility values of tasks that generate a large number of tokens (i.e., those with longer execution cycles), such long tasks can be prioritized for exclusion during subsequent task selection. This design mimics the SJF scheduling strategy, effectively avoiding head-of-line blocking caused by prolonged resource occupancy. Conversely, if continuous execution of specific tasks needs to be ensured to prevent mid-process preemption, the utility value of those tasks can be appropriately increased to give them a competitive advantage in scheduling, thereby reducing the likelihood of interruption.

\begin{algorithm}[htp]
	\caption{SLO-Driven online Scheduling: SLICE-online}\label{alg:SLICE_online}
	\begin{algorithmic}[1]
		\STATE \textbf{Input:} Task set $\mathbb{N}=\{0,1,2,...,N\}$ for receiving new tasks; TPOT value set $\mathbb{T}=\{T_{TPOT}^{0},T_{TPOT}^{1},...,T_{TPOT}^{N}\}$; Utility value set $\mathbb{U}=\{U_0,U_1,...,U_N\}$; For any task $i\in \mathbb{N}$: TPOT requirement $T_{TPOT}^{i}$, Utility value $U_i$ ; Token buffer $tokenBuf$ for streaming response; Event queue $eventQ$
		\STATE \textbf{Output:} Tasks not yet executed
		\STATE $tokenBuf,eventQ\gets \emptyset,\emptyset$;
		\STATE \textbf{while} True \textbf{do}\label{task_exe_event_begin}
		\\ \hspace{0.5cm} $\triangleright$Start a schedule
		\STATE \hspace{0.5cm} Start thread SLICEoffline$(\mathbb{N},\mathbb{T},\mathbb{U},tokenBuf,eventQ)$
		\STATE \hspace{0.5cm} \textbf{while} True \textbf{do}
		\STATE \hspace{0.5cm}\hspace{0.5cm} \textbf{if} a task is arrived or completed \textbf{do} 
		\STATE \hspace{0.5cm}\hspace{0.5cm}\hspace{0.5cm}$eventQ.put(reshedule\, message)$
		\STATE \hspace{0.5cm}\hspace{0.5cm}\hspace{0.5cm}\textbf{break}
		\STATE \hspace{0.5cm}\hspace{0.5cm} \textbf{end if} 
		\STATE \hspace{0.5cm}\hspace{0.5cm} \textbf{if} other reschedule event is arrived \textbf{do}
		\STATE \hspace{0.5cm}\hspace{0.5cm}\hspace{0.5cm}$eventQ.put(reshedule\, message)$
		\STATE \hspace{0.5cm}\hspace{0.5cm}\hspace{0.5cm}\textbf{break}
		\STATE \hspace{0.5cm}\hspace{0.5cm}\textbf{end if}
		\STATE \hspace{0.5cm} \textbf{end while}
		\STATE \hspace{0.5cm}$\mathbb{N}\gets$\textbf{waiting} SLICEoffline
		\\\hspace{0.5cm}$\triangleright$ Used for controlling task preemption
		\STATE \hspace{0.5cm}$\mathbb{U}\gets$\textsc{UtilityAdaptor}$(\mathbb{U},\mathbb{T},\mathbb{N})$\label{task_utility_adapt}
		\STATE \textbf{end while}\label{task_exe_event_end}
		\STATE \textbf{return} $\mathbb{N}$\quad$\triangleright$ Return tasks not yet executed 
	\end{algorithmic}
\end{algorithm}

\section{System Implementation}\label{sec_system}
\begin{figure}[htp]
	\centering
	\includegraphics[width=0.99\linewidth]{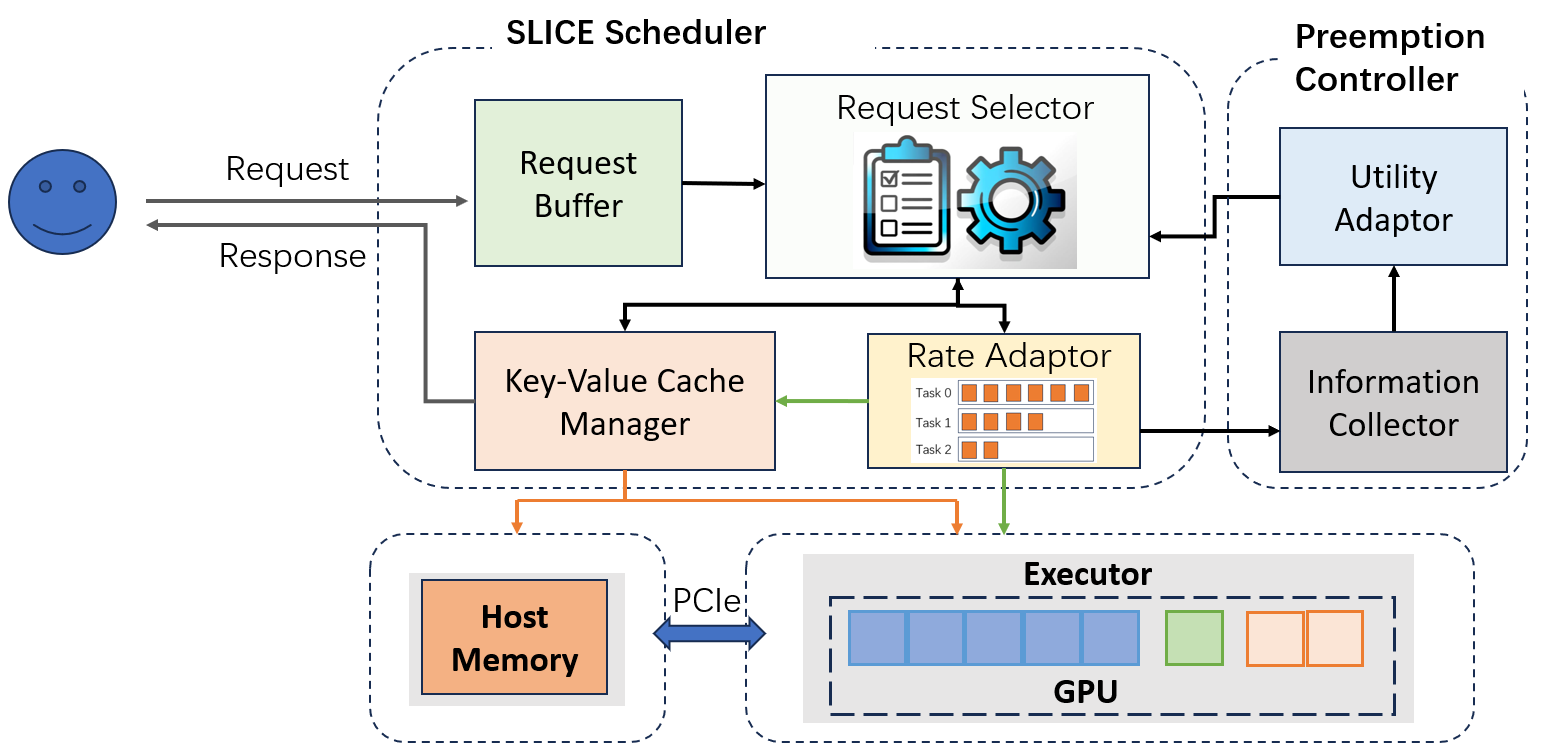}
	\caption{Overview of SLICE implementation. The system consists of the SLICE scheduler and the preemption controller.}
	\label{fig_system_overview}
\end{figure}

The online SLICE scheduling algorithm is designed to be universal, exhibiting no dependency on specific inference systems and enabling seamless integration with mainstream large language model (LLM) inference platforms. Among these, FastLLM is a lightweight LLM inference framework developed entirely in C++, requiring no third-party plugin support and specifically optimized for low-resource environments on edge computing devices. We implemented the SLICE online algorithm using approximately 1,000 lines of C++ code and integrated it into the FastLLM platform. The system consists of two components: \textbf{the SLICE Scheduler} and \textbf{the Preemption Controller}, as illustrated in Fig. \ref{fig_system_overview}. 

{\bf SLICE Scheduler.} User requests for LLM first enter the request buffer of the SLICE Scheduler to await scheduling. Upon the occurrence of events such as task arrival, departure, or other scheduling triggers, the SLICE Scheduler selects a batch tasks from the buffer according to Algorithm \ref{alg:task_select} and forwards them to the SLO-aware rate allocator according to Algorithm \ref{alg:rate_Allocation}. The rate allocator then assigns each task a decoding rate that no less than its SLO requirement.

{\bf Preemption Controller.} When a scheduling event occurs, the Preemption Controller dynamically adjusts the utility values of currently executing tasks, thereby enabling a flexible multi-objective preemptive scheduling mechanism as described in Section \ref{subsec:off2online}.

\section{Evaluations}\label{sec_evaluation}
\subsection{Experimental Settings}
\textbf{Testbed.} This evaluation is conducted on a test platform equipped with an NVIDIA RTX 4060Ti (16GB GDDR6), where the GPU is connected to the host system via a PCIe 3.0 x16 interface. The host machine is configured with 22GB of system memory and an Intel Xeon E5-2690 v4 processor.

\textbf{LLM models.} The evaluation utilizes the open source LLM ChatGLM2-6B-INT4, selected for its exceptional deployment flexibility. As an INT4-quantized model, it requires only ~6GB GPU memory and is widely compatible with consumer-grade edge computing hardware, striking a balance between efficiency and accessibility. This design ensures fast inference and streamlined experimentation while significantly improving research feasibility and reproducibility.

\textbf{Workloads.} To comprehensively evaluate the system's performance under varying load intensities, this study designs a comprehensive workload scheme incorporating multiple types of tasks. Specifically, the experiment configures two task types with differentiated SLO requirements:
\begin{enumerate}
	\item Real-time tasks, such as machine control and navigation planning \cite{ling2024timelyllm}, demand strict adherence to response rates (i.e., at least 20 tokens/s) to ensure critical tasks are completed within deadlines (i.e., 1.5s).  
	\item Non-real-time tasks, such as voice chat tasks (8 tokens/s) %require generation rates matching natural speech speed to maintain conversational fluency. 
	and text-based Q\&A tasks (10 tokens/s),  generate content at a pace close to the average human reading speed to ensure a smooth interactive experience.
\end{enumerate}
In the workload design, task arrivals are simulated using a Poisson process, with ten incrementally increasing request arrival rates (ranging from 0.1 to 7.0 tasks/sec). This approach systematically examines the system's performance evolution from idle to overloaded states while assessing SLO compliance across different task types. Such a refined experimental setup enables a thorough analysis of the system's performance characteristics and potential bottlenecks under varying load conditions.

\textbf{Baselines.} To comprehensively validate the performance advantages of the proposed SLICE scheduling in this study, we have selected two state-of-the-art baseline strategy for comparative analysis:
\begin{enumerate}
	\item ORCA: ORCA \cite{yu2022orca} first proposed Dynamic Batching method, an innovative approach that has now become the standard scheduling strategy in mainstream inference frameworks. Leading systems such as FastLLM, FasterTransformer, and vLLM have adopted it as their default implementation, making it a crucial benchmark for evaluating algorithmic improvements.  
	\item FastServe: FastServe \cite{wu2023fast} innovatively utilizes task generation length information to achieve dynamic priority adjustment, effectively mitigating head-of-line blocking through its iterative preemption mechanism . %The scheme has been adopted by advanced serving frameworks like FastServe \cite{wu2023fast} and has demonstrated significant performance advantages in practical scenarios.
\end{enumerate}
Through this dual comparison, we can not only verify our solution's improvements over conventional methods but also highlight its innovative value in addressing complex scheduling challenges.

\textbf{Metrics.} This study employs three core metrics - TTFT attainment, TPOT attainment, and SLO attainment - to systematically evaluate the comprehensive performance of scheduling strategies. Specifically, for real-time tasks, their SLOs are achieved if and only if their completion time does not exceed the deadline requirement. While for non-real-time tasks, their SLOs are considered fulfilled only when both the TTFT SLO and TPOT SLO are simultaneously satisfied.

\subsection{Static performance}

This experiment designed three types of experimental tasks with differentiated TPOT requirements, each containing multiple independent requests, totaling nine experimental tasks. The detailed settings are provided in Table \ref{table_static_tpot}. Based on this experimental setup, the effectiveness of the SLICE algorithm in dynamically allocating resources for different TPOT requests was systematically evaluated.

\begin{figure}[htp]
	\centering
	\includegraphics[width=0.9\linewidth]{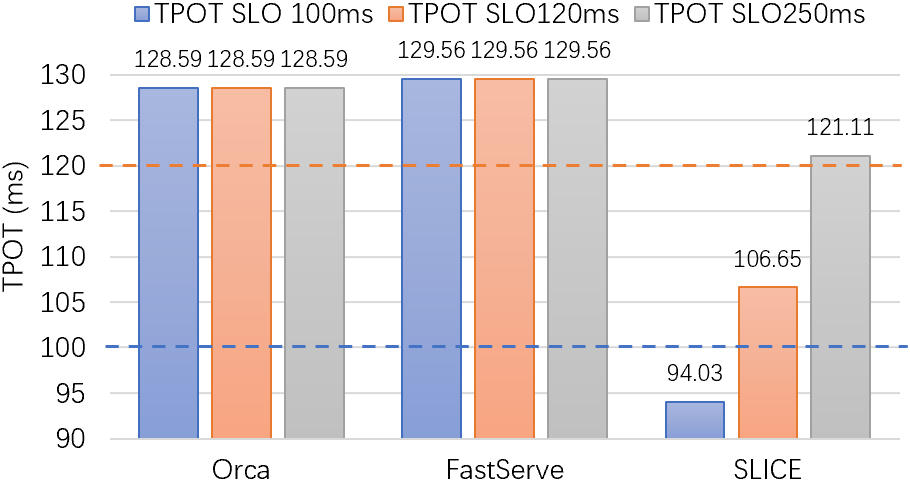}
	\caption{TPOT Statistics of Different Task Types Under Three Scheduling Strategies. SLICE can allocate corresponding rates based on the TPOT requirements of each task to meet the diverse TPOT demands of various tasks. In contrast, Orca and FastServe adopt a uniform decoding rate strategy, assigning the same decoding rate (i.e., TPOT value) to all tasks without considering the heterogeneous requirements among them.}
	\label{fig_static_tpot}
\end{figure}

\begin{table*}[htp]
	\centering
	\caption{TPOT statistics of different task type under three scheduling}
	\label{table_static_tpot}
	\begin{tabular}{ c|c|c| c | c|c |c|c}
		\hline
		Strategy&Task Type&Task number&TPOT SLO&Actual TPOT&Decoding rate&Is TPOT satisfied&SLO attainment\\
		\hline
		\multirow{3}{*}{Orca}
		&Task A&3&100ms&128.59ms&7.78  tokens/s&No&\\
		\cline{2-7}
		&Task B&4&120ms&128.59ms&7.78  tokens/s&No&22\%\\
		\cline{2-7}
		&Task C&2&250ms&128.59ms&7.78 tokens/s&Yes&\\
		\hline
		\multirow{3}{*}{FastServe}
		&Task A&3&100ms&129.56ms&7.72  tokens/s&No&\\
		\cline{2-7}
		&Task B&4&120ms&129.56ms&7.72  tokens/s&No&22\%\\
		\cline{2-7}
		&Task C&2&250ms&129.56ms&7.72 tokens/s&Yes&\\
		\hline
		\multirow{3}{*}{\textbf{SLICE}}
		&Task A&3&100ms&\textbf{94.03ms}&10.63  tokens/s&\textbf{Yes}&\\
		\cline{2-7}
		&Task B&4&120ms&\textbf{106.65ms}&9.37  tokens/s&\textbf{Yes}&\textbf{100\%}\\
		\cline{2-7}
		&Task C&2&250ms&\textbf{121.11ms}&8.26 tokens/s&\textbf{Yes}&\\
		\hline
	\end{tabular}	
\end{table*}

As shown in Table 1, the experimental configuration of this experiment includes three types of heterogeneous tasks (A, B, and C), with their TPOT SLOs set to 100 ms, 120 ms, and 250 ms, respectively. Specifically, Task Type A consists of 3 independent request instances, Task Type B comprises 4 independent request instances, and Task Type C contains 2 independent request instances. The experiment employs three scheduling strategies for comparison: the proposed SLICE algorithm, the current state-of-the-art method Orca, and FastServe. Through experimental measurements, the actual TPOT values obtained by each task type during runtime are statistically analyzed (with specific numerical distributions illustrated in Fig. \ref{fig_static_tpot}).

From the perspective of rate allocation, SLICE can dynamically assign transmission rates that precisely match the TPOT demand characteristics of each individual task, thereby ensuring that the actual TPOT values of all task types align with their preset SLOs. In contrast, during practical operation, both the Orca and FastServe scheduling strategies allocate identical TPOT values to all tasks. This means they employ a uniform decoding rate across tasks, as shown in Fig. \ref{fig_static_tpot}. The fundamental reason for this phenomenon lies in the fact that Orca and FastServe merge all pending tasks into a single consolidated batch for LLM's forward pass during each decoding cycle. This processing approach rigidly enforces consistent decoding rates across all tasks (i.e., identical TPOT values for every task), consequently eliminating their ability to accommodate the heterogeneous demand characteristics among different tasks.

From the perspective of SLO attainment, the SLICE algorithm achieved a 100\% SLO attainment. In contrast, Orca and FastServe exhibited an SLO attainment of merely 22\% (as shown in Table \ref{table_static_tpot}). The fundamental cause of this significant discrepancy lies in the large batch processing scale employed by Orca and FastServe during each decoding cycle, which leads to a substantial increase in overall decoding latency. This processing approach renders high TPOT demand tasks (such as Type A and Type B) difficult to meet their stringent latency requirements, while only low TPOT demand tasks (such as Type C) with relatively lenient requirements can be satisfied (see Table \ref{table_static_tpot}).

\subsection{Dynamic performance}
In this section's experiments, we construct an scenario with dynamic task arrivals and departures. And the task arrival rate is set to 1, a load condition that has been tested to precisely saturate the performance capacity of the experimental GPU. Considering that real-time tasks typically dominate in edge computing scenarios, we configured a 7:3 ratio between real-time and non-real-time tasks. Under this setup, we conducted a comparative evaluation of ORCA, FastServe, and SLICE strategies using four key metrics: SLO attainment, TTFT SLO attainment, TPOT SLO attainment, and average task completion time.

\begin{figure}[htp]
	\centering
	\includegraphics[width=0.9\linewidth]{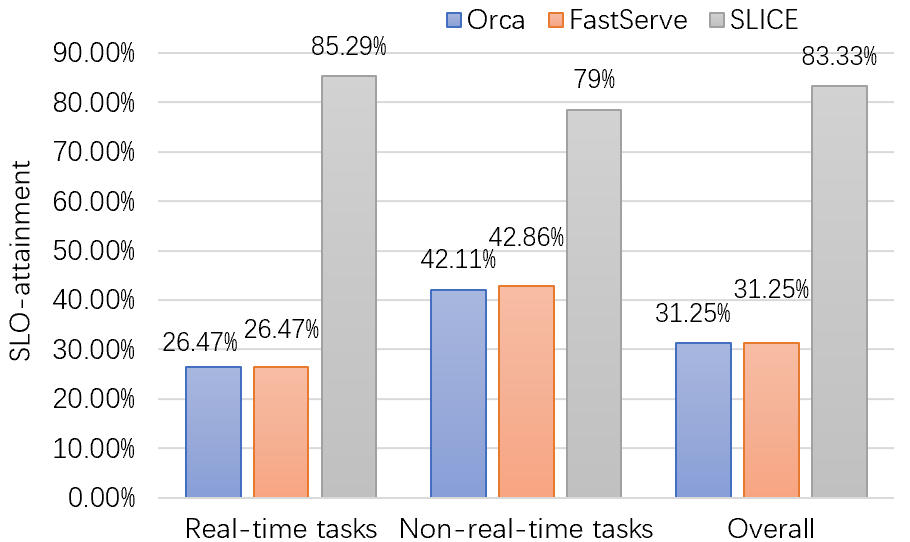}
	\caption{SLO attainment comparison of real-time vs. non-real-time tasks for the three strategies. Our SLICE demonstrates significant performance superiority over state-of-the-art methods Orca and FastServe in SLO attainment metrics across all task types.}
	\label{fig_slo}
\end{figure}

\textbf{Performance of SLO attainment.} As illustrated in Fig. \ref{fig_slo}, the SLICE exhibits substantial improvements in SLO attainment metrics. In overall performance, SLICE achieves a SLO attainment of 83.33\%, representing a 2.67× improvement over baseline methods ORCA and FastServe (both at 31.25\% SLO attainment). For real-time task processing, SLICE attains an exceptional SLO attainment of 85.29\%, achieving a 3.23× performance enhancement. In non-real-time scenarios, it maintains a robust 78.15\% attainment rate with an average 1.92× performance gain. %To thoroughly investigate the underlying mechanisms behind the superior performance of the SLICE algorithm compared to the two baseline algorithms (ORCA and FastServe), this study further selects and systematically presents three key metrics—the TPOT SLO attainment, TTFT SLO attainment, and deadline SLO attainment—for in-depth analysis.

\begin{figure}[htp]
	\centering
	\includegraphics[width=0.9\linewidth]{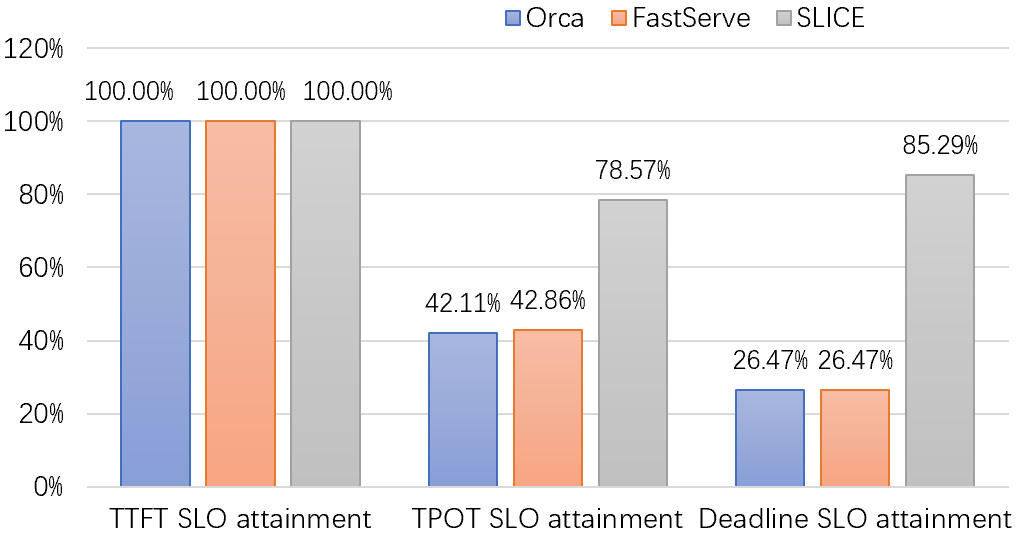}
	\caption{Comparison of TPOT SLO attainment, TTFT SLO attainment, and deadline SLO attainment for the three strategies. In non-real-time task scenarios, all three strategies achieved 100\% TTFT SLO attainment. However, regarding TPOT metrics, both ORCA and FastServe demonstrated relatively low SLO attainment rates, while the SLICE achieved a remarkable 78.5\% optimization. For real-time tasks, ORCA and FastServe could only ensure approximately 26\% of tasks were completed before their deadlines. In contrast, SLICE significantly improved this ratio to nearly 85\%, demonstrating superior real-time task processing capabilities.}
	\label{fig_detailed_slo}
\end{figure}

As we can see in Fig. \ref{fig_slo}, ORCA and FastServe exhibit comparable performance characteristics. The fundamental reason lies in their inherent limitation in edge computing scenarios where the task arrival rate consistently fails to reach their predefined maximum batch processing capacity. Whenever tasks enter the system, they are immediately incorporated into the current batch for unified decoding, resulting in identical task scheduling strategies between the two algorithms. When the number of tasks within a batch accumulates to a specific threshold, the cumulative effect of decoding latency directly triggers a significant increase in TPOT. For non-real-time tasks, this latency escalation elevates the TPOT SLO violation probability across all task types despite maintaining 100\% TTFT SLO attainment, as shown in Fig. \ref{fig_detailed_slo}. For real-time tasks, the immediate consequence is substantial prolongation of task completion time, causing tasks to miss their predefined deadlines and ultimately leading to a marked rise in real-time task SLO violation rates. In contrast, the SLICE implements selective batch scheduling based on each task's specific SLO requirements, while flexibly allocating differentiated decoding rates to tasks within the same batch, thereby achieving more precise performance optimization.

\begin{figure}[htp]
	\centering
	\includegraphics[width=0.9\linewidth]{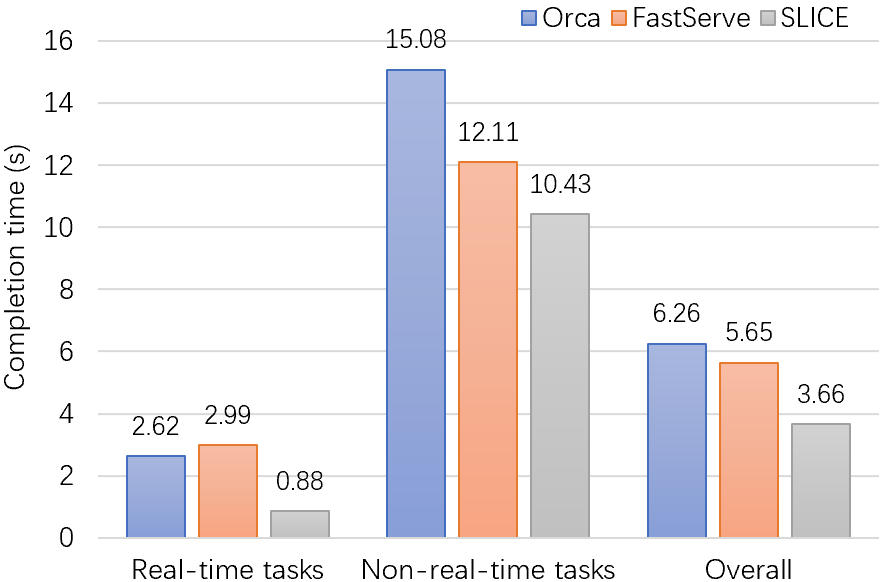}
	\caption{Completion time comparison of real-time vs. non-real-time tasks for the three algorithms. Our SLICE algorithm significantly outperforms the comparative algorithms ORCA and FastServe.}
	\label{fig_jct}
\end{figure}

\textbf{Performance of task completion time.} In terms of task completion time, the SLICE significantly outperforms the comparative methods ORCA and FastServe, as illustrated in Fig. \ref{fig_jct}. For real-time tasks, it demonstrates an average completion time advantage of 2.9x over ORCA and 3.4x over FastServe. Regarding non-real-time tasks, the average completion time improvements are 1.4x and 1.2x compared to ORCA and FastServe, respectively. Across all task types, SLICE achieves notable optimizations with approximately 1.7x and 1.5x faster average completion times than ORCA and FastServe, respectively.

\subsection{Performance varying with task ratio}
In this section's experiments, we fix the task arrival rate at 1 and adjust the proportion of real-time tasks in the total workload. Under this experimental setup, we compare the performance of Orca, FastServe and the proposed SLICE in terms of SLO attainment.
\begin{figure*}[htp]
	\centering
    \subfloat[Real-time tasks' SLO attainment]{\includegraphics[width=0.33\linewidth]{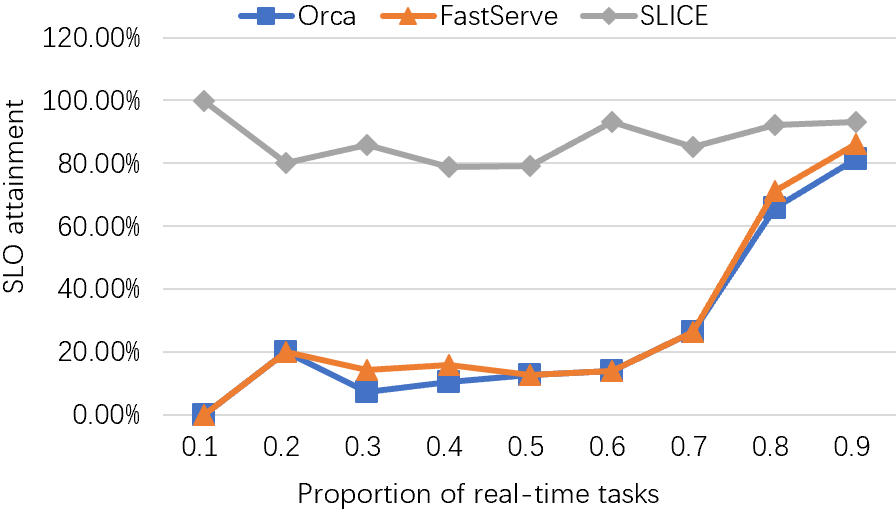}%
	\label{fig_slo_ratio_a}}
    \hfil
    \subfloat[Non-real-time tasks' SLO attainment]{\includegraphics[width=0.33\linewidth]{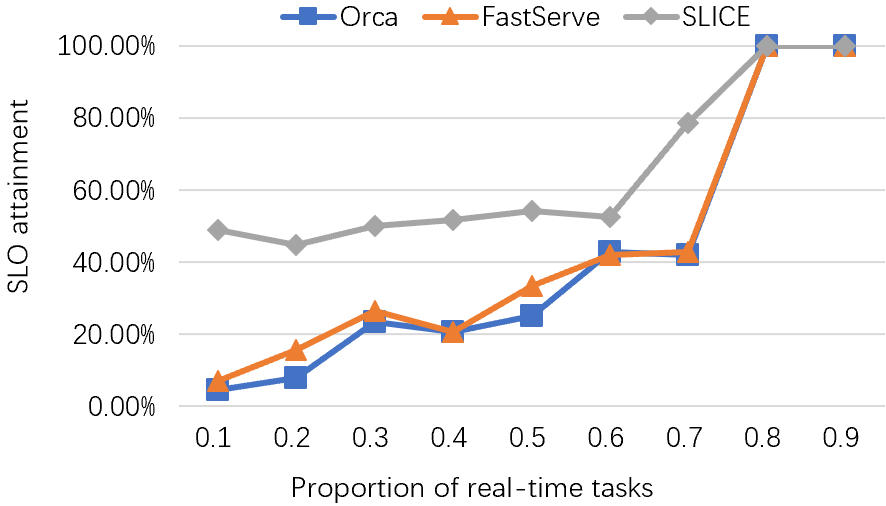}%
	\label{fig_slo_ratio_b}}
    \hfil
    \subfloat[Overall SLO attainment]{\includegraphics[width=0.33\linewidth]{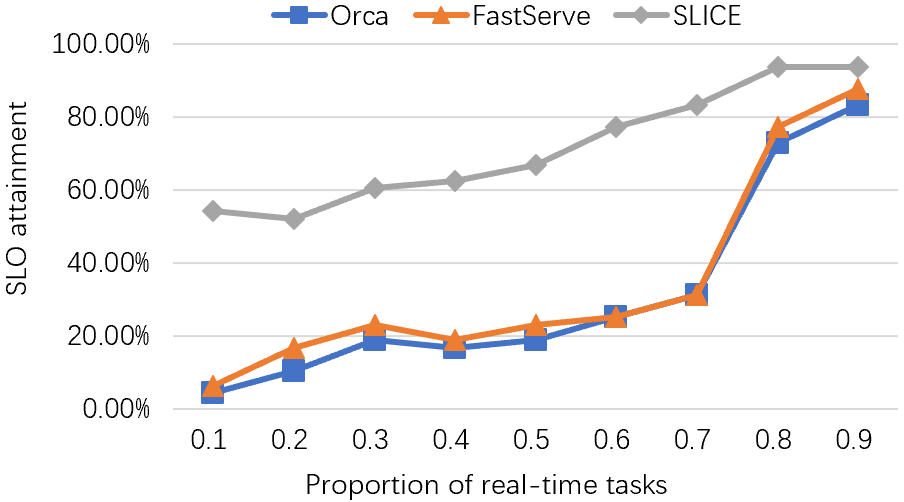}%
	\label{fig_slo_ratio_c}}
	\caption{SLO attainment across different task types. (a) SLICE consistently maintains an SLO achievement rate above 80\% for real-time tasks; in contrast, Orca and FastServe only achieve an SLO rate of approximately 10\% when the proportion of real-time tasks is below 70\%. (b) SLICE consistently achieves the highest SLO attainment for non-real-time tasks across varying task ratios compared to both Orca and FastServe. Notably, when real-time tasks account for just 10\% of the workload, SLICE attains an SLO attainment 10.5× higher than that of the other two methods. (c)Overall, across all task mix scenarios, SLICE demonstrates consistently superior SLO achievement rates compared to both Orca and FastServe, with performance advantages reaching up to 13× in certain configurations.}
    \label{fig_slo_ratio}
\end{figure*}

As we can see in Fig. \ref{fig_slo_ratio_c}, SLICE demonstrates a significantly superior SLO achievement rate compared to Orca and FastServe across all task scenarios, with its performance advantage reaching up to 13×. Specifically, for real-time task processing, SLICE consistently maintains a stable SLO achievement rate exceeding 80\%. In contrast, when the proportion of real-time tasks falls below 70\%, Orca and FastServe only achieve a comparatively low SLO attainment of approximately 10\% (see Fig. \ref{fig_slo_ratio_a}). For non-real-time tasks, SLICE still maintains the best SLO attainment performance. When real-time tasks account for 10\% of the workload, the SLO attainment of SLICE is 10.5× higher than the other two methods (see Fig. \ref{fig_slo_ratio_b}).

When the proportion of real-time tasks decreases, the SLO attainment of both SLICE and the comparative methods (Orca/FastServe) exhibit a declining trend. This phenomenon stems from the fundamental differences between real-time tasks and long-duration non-real-time tasks. Real-time tasks typically consist of short-duration operations requiring rapid responses (such as machine control commands), whereas non-real-time tasks generally feature longer execution cycles. As the percentage of real-time tasks diminishes, the proportion of long-duration tasks in the system correspondingly increases, leading to a significant escalation in overall system load. When this load pressure persistently exceeds the hardware performance limits of the GPU, all three scheduling strategies demonstrate a degradation in their SLO achievement rates.

Notably, SLICE maintains a relative advantage in this scenario, demonstrating significantly smaller performance degradation compared to the baseline algorithms, as illustrated in \ref{fig_slo_ratio}. This advantage stems from fundamental differences in their scheduling mechanisms. Orca and FastServe employ a coarse-grained batch processing approach, where all arriving tasks are aggregated into a single batch for synchronous decoding in our experiment settings. When the batch size exceeds 8 tasks, the decoding latency surpasses the 100ms threshold, resulting in severe TPOT violations. In contrast, SLICE innovatively adopts a dynamic rate allocation strategy. During decoding, it adaptively adjusts the decoding rates of individual tasks within the batch based on their specific requirements, thereby ensuring more tasks' TPOT requirement. When computational resources become insufficient to meet the TPOT demands of additional tasks, the system proactively halts batch expansion. By effectively controlling batch size and thus decoding latency in this manner, SLICE maintains a baseline TOPT satisfaction rate, thereby preserving relatively reliable SLO attainment even under overload conditions.

\subsection{Performance varying with workloads}
In this section's experiments, we set the ratio of real-time tasks to non-real-time tasks at 7:3, and simulate experimental environments under varying workload conditions by progressively adjusting the task arrival rate (from 0.1 to 7). The higher the task arrival rate, the heavier the system workload. For each arrival rate, we evaluate the performance of the three scheduling methods in terms of SLO attainment.

\begin{figure*}[htp]
	\centering
	\subfloat[Real-time tasks' SLO attainment]{\includegraphics[width=0.33\linewidth]{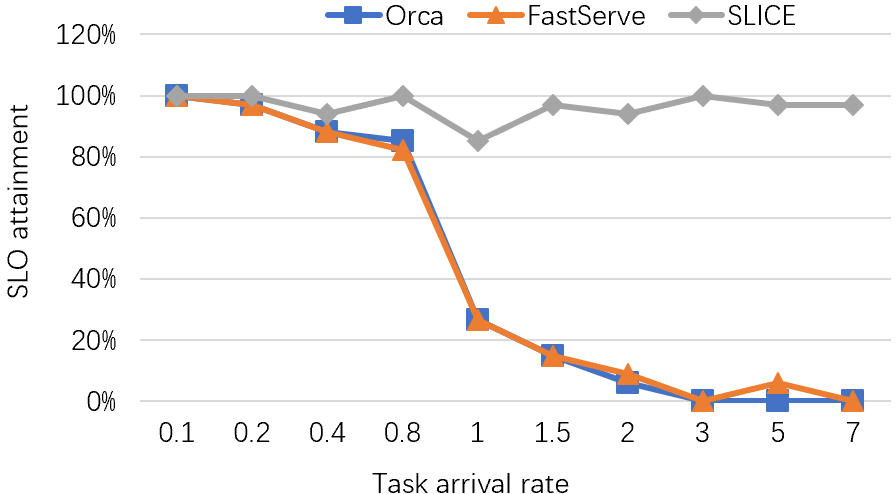}%
		\label{fig_slo_workload_a}}
	\hfil
	\subfloat[Non-real-time tasks' SLO attainment]{\includegraphics[width=0.33\linewidth]{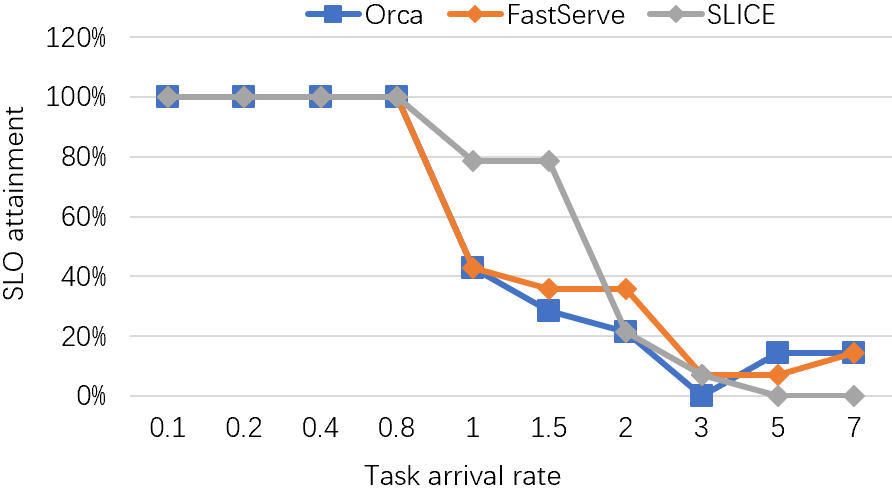}%
		\label{fig_slo_workload_b}}
	\hfil
	\subfloat[Overall SLO attainment]{\includegraphics[width=0.33\linewidth]{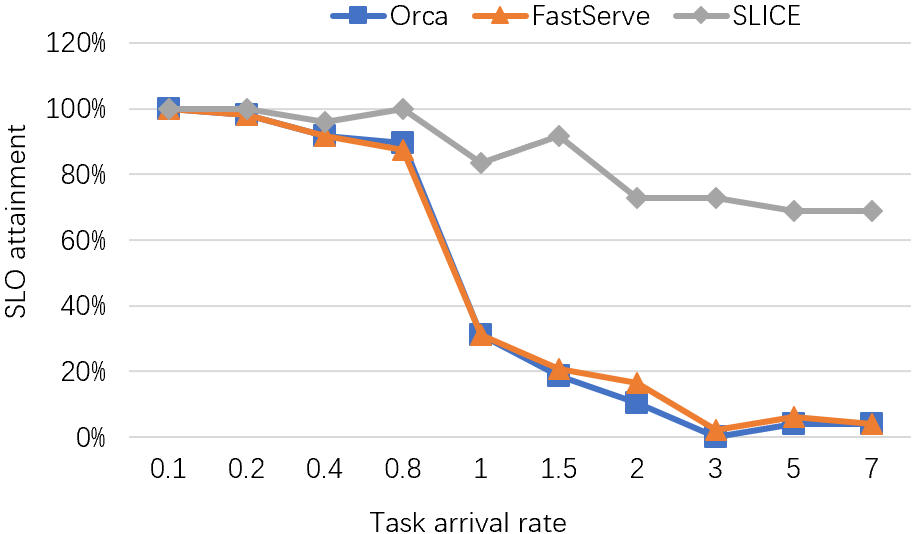}%
		\label{fig_slo_workload_c}}
	\caption{SLO attainment across different workloads. (a) SLICE consistently maintains an SLO attainment close to 100\% in real-time task scenarios. In contrast, when the task arrival rate exceeds 1.5, the SLO attainment of Orca and FastServe for real-time tasks drop to nearly 0. (b) When the task arrival rate is below 2, SLICE achieves a higher SLO attainment than Orca and FastServe in non-real-time task scenarios. However, when the arrival rate exceeds 2, SLICE and both baseline methods maintain relatively low SLO attainment. (c) Overall, SLICE demonstrates consistently superior SLO attainment across varying workload conditions, significantly outperforming both Orca and FastServe. Notably, at an arrival rate of 3, SLICE achieves an SLO attainment 35x higher than that of the other two methods.}
	\label{fig_slo_workload}
\end{figure*}

Overall, SLICE demonstrates exceptional SLO attainment performance across various workload conditions, consistently outperforming Orca and FastServe with a maximum performance advantage of up to 35×. When the task arrival rate exceeds 0.8, the SLO achievement rates of Orca and FastServe decline sharply, whereas SLICE experiences a moderate decrease but still maintains a relatively high SLO achievement rate, as shown in Fig. \ref{fig_slo_workload_c}. The root cause of this phenomenon lies in Orca and FastServe's excessively large batch sizes under high arrival rates. Moreover, all tasks within the batch must participate decoding every iteration, resulting in prohibitively high decoding latency. The tokens' output TPOT far exceeds the SLO requirement, leading to severe TPOT SLO violations. In contrast, SLICE precisely regulates the decoding rate of scheduled tasks based on their individual TPOT requirements, and thus can  allocate limited computational resources across more tasks. This dynamic resource management strategy achieves superior SLO attainment.

In real-time task scenarios, SLICE maintains a near 100\% SLO attainment, whereas Orca and FastServe experience a sharp decline in SLO attainment, dropping to nearly 0\% when the task arrival rate exceeds 1.5 (see Fig. \ref{fig_slo_workload_a}). This is because SLICE assigns higher utility values to real-time tasks, enabling priority scheduling of real-time tasks to ensure their service guarantees.

For non-real-time tasks, SLICE significantly outperforms the comparative methods when the arrival rate is below 2, while all methods including SLICE exhibit relatively low SLO attainment when the arrival rate surpasses 2 (see Fig. \ref{fig_slo_workload_b}). This occurs because, under high arrival rates with limited computational resources, SLICE prioritizes the allocation of resources to real-time tasks. As a result, non-real-time tasks receive comparatively fewer computational resources, leading to a higher rate of SLO violations for non-real-time tasks. Although the SLO attainment for non-real-time tasks is relatively low under high workload conditions, SLICE guarantees nearly 100\% SLO attainment for real-time tasks. Remarkably, even when the task arrival rate exceeds 1.5, SLICE maintains an overall SLO attainment of approximately 80\% (see Fig. \ref{fig_slo_workload_c}).

\section{Conclusion} \label{sec_conclusion}
The integration of LLMs into edge computing systems marks a pivotal advancement in embodied intelligence, enabling real-time applications across humanoid robots, autonomous vehicles, and other latency-sensitive scenarios. However, the inherent diversity of SLOs, such as various constraints on end-to-end latency, TTFT and TPOT, poses critical challenges for existing scheduling frameworks that prioritize throughput over differentiated SLO requirements.

This paper addresses this research gap by proposing SLICE, a SLO-driven scheduling framework that prioritizes real-time task execution while maximizing the SLO attainment of non-real-time tasks. The SLICE dynamically adjusts decoding rates via a decode-mask matrix, allocating fewer resources to tasks with lower demands under strict SLO constraints, thereby enabling limited resources to support more concurrent tasks. Through rigorous experimentation, SLICE achieves up to 35× improvement in SLO attainment under heavy workloads and a 3.4× advantage in average completion time compared to state-of-the-art benchmarks Orca and FastServe. These results underscore the necessity of customizing LLM scheduling mechanisms for edge-specific SLOs to ensure reliable responses.

\ifCLASSOPTIONcaptionsoff
\newpage
\fi



\bibliographystyle{IEEEtran}
\bibliography{bibliography}
%\begin{thebibliography}{1}


%\end{thebibliography}


%\vfill

\end{document}